\documentclass[longauth,traditabstract]{aa} 
\usepackage{graphicx}
\usepackage{lscape}
\usepackage{txfonts}
\usepackage{natbib}
\usepackage{url}
\usepackage[T1]{fontenc}

\usepackage[ps2pdf,pdfpagemode=UseThumbs,colorlinks=true,bookmarks=true]{hyperref}
\hypersetup{citecolor={blue},linkcolor={blue},urlcolor={blue}}

\newcommand{\fuden}{10$^{-17}$ erg~s$^{-1}$~cm$^{-2}$~arcsec$^{-1}$}
\newcommand{\fuDEN}{10$^{-16}$ erg~s$^{-1}$~cm$^{-2}$~arcsec$^{-1}$}

\newcommand{\FunitsA}{10$^{-16}$ erg~s$^{-1}$~cm$^{-2}$~arcsec$^{-2}$}
\newcommand{\degree}{\ensuremath{^\circ}}

\newcommand{\hh}{\ion{H}{ii}~}

\newcommand{\nii}{[\ion{N}{ii}]}

\newcommand{\oii}{[\ion{O}{ii}]}
\newcommand{\oiii}{[\ion{O}{iii}]}
\newcommand{\sii}{[\ion{S}{ii}]}

\newcommand{\lam}{$\lambda$}

\newcommand{\ha}{H$\alpha$} 
\newcommand{\hb}{H$\beta$} 
\newcommand{\hg}{H$\gamma$} 
\newcommand{\flux}{erg\,s$^{-1}$\,cm$^{-2}$}

\newcommand{\mz}{{\small $\mathcal{M}$-Z}}
\newcommand{\Sz}{{\small {\large $\Sigma$}-Z}}

\DeclareRobustCommand{\ion}[2]{%
\relax\ifmmode
\ifx\testbx\f@series
{\mathbf{#1\,\mathsc{#2}}}\else
{\mathrm{#1\,\mathsc{#2}}}\fi
\else\textup{#1\,{\mdseries\textsc{#2}}}%
\fi}

\newcommand{\HII}{\ion{H}{ii}~}

\begin{document}
   \title{A characteristic oxygen abundance gradient in galaxy disks unveiled with CALIFA}

   \author{
     S.\,F. S\'anchez\inst{\ref{iaa},\ref{caha},\ref{unam}}
     \and
     F.\,F. Rosales-Ortega\inst{\ref{inaoe}}
     \and
     J. Iglesias-P\'aramo\inst{\ref{iaa},\ref{caha}}
     \and
     M. Moll\'a\inst{\ref{ciemat}}
     \and
     J. Barrera-Ballesteros \inst{\ref{iac}}     
     \and
     R.\,A. Marino\inst{\ref{ucm}}
     \and
     E. P\'erez \inst{\ref{iaa}}
     \and
     P.\, S\'anchez-Blazquez\inst{\ref{uam}}
     \and
     R. Gonz\'alez Delgado\inst{\ref{iaa}}
     \and
     R. Cid Fernandes\inst{\ref{flora}}
     \and
      A. de Lorenzo-C\'aceres\inst{\ref{lalag},\ref{StAnd}}
     \and
      J. Mendez-Abreu \inst{\ref{iac},\ref{lalag},\ref{StAnd}}
     \and
     L. Galbany\inst{\ref{centra}}
     \and
     J. Falcon-Barroso \inst{\ref{iac}}     
     \and
     D.\,Miralles-Caballero\inst{\ref{uam}}
     \and
     B.\, Husemann\inst{\ref{aip}}
     \and
     R. Garc\'\i a-Benito \inst{\ref{iaa}}
     \and
     D. Mast\inst{\ref{caha},\ref{iaa}}
     \and
     C.\,J.\, Walcher\inst{\ref{aip}}
     \and
     A. Gil de Paz\inst{\ref{ucm}}
     \and
      B. Garc\'\i a-Lorenzo\inst{\ref{iac}}     
     \and
     B. Jungwiert\inst{\ref{prague}}
     \and
     J.\,M. V\'\i lchez\inst{\ref{iaa}}
     \and
     Lucie J\'{i}lkov\'{a}\inst{\ref{brno}}
     \and
     M. Lyubenova \inst{\ref{mpia}}
     \and
     C. Cortijo-Ferrero\inst{\ref{iaa}}
     \and
     A.\,I.\, D\'\i az\inst{\ref{uam}}
     \and
     L.\, Wisotzki\inst{\ref{aip}}
\and
     I.\, M\'arquez\inst{\ref{iaa}}
     \and
     J. Bland-Hawthorn \inst{\ref{sydney}}
      \and
     S. Ellis\inst{\ref{sydney},\ref{aao}}
     \and
     G. van de Ven \inst{\ref{mpia}}
     \and
     K. Jahnke \inst{\ref{mpia}}
     \and     
      P. Papaderos \inst{\ref{porto}}
     \and
      J. M. Gomes\inst{\ref{porto}}
     \and
      M. A. Mendoza\inst{\ref{iaa}}
     \and
     \'A. R. L\'opez-S\'anchez\inst{\ref{sydney},\ref{aao}}
      \and
     { The CALIFA collaboration}
          }

   \institute{
        \label{iaa}Instituto de Astrof\'{\i}sica de Andaluc\'{\i}a (CSIC), Glorieta de la Astronom\'\i a s/n, Aptdo. 3004, E18080-Granada, Spain\\ \email{sanchez@iaa.es}.
        \and
        \label{caha}Centro Astron\'omico Hispano Alem\'an, Calar Alto, (CSIC-MPG), C/Jes\'{u}s Durb\'{a}n Rem\'{o}n 2-2, E-04004 Almer\'{\i}a, Spain.
        \and
        \label{unam}Instituto de Astronom\'\i a,Universidad Nacional Auton\'oma de Mexico, A.P. 70-264, 04510, M\'exico,D.F.
       \and
       \label{inaoe} Instituto Nacional de Astrof{\'i}sica, {\'O}ptica y Electr{\'o}nica, Luis E. Erro 1, 72840 Tonantzintla, Puebla, Mexico
\and
        \label{ciemat}Departamento de Investigaci\'on B\'asica, CIEMAT, Avda. Complutense 40 E-28040 Madrid, Spain.
\and
\label{iac}Instituto de Astrof\'\i sica de Canarias (IAC), E-38205 La Laguna, Tenerife, Spain 
       \and
\label{ucm}CEI Campus Moncloa, UCM-UPM, Departamento de Astrof\'{i}sica y CC$.$ de la Atm\'{o}sfera, Facultad de CC$.$ F\'{i}sicas, Universidad Complutense de Madrid, Avda.\,Complutense s/n, 28040 Madrid, Spain.
       \and
       \label{uam}Departamento de F\'isica Te\'orica, Universidad Aut\'onoma de Madrid, 28049 Madrid, Spain.
\and
\label{flora}Departamento de F\'{\i}sica, Universidade Federal de Santa Catarina, P.O. Box 476, 88040-900, Florian\'opolis, SC, Brazil
\and
\label{lalag}Depto. Astrof\'\i sica, Universidad de La Laguna (ULL), E-38206 La Laguna, Tenerife, Spain
\and
\label{StAnd}School of Physics and Astronomy, University of St Andrews, North Haugh, St Andrews, KY16 9SS, U.K. (SUPA)
\and
\label{centra}CENTRA - Instituto Superior Tecnico, Av. Rovisco Pais, 1, 1049-001 Lisbon, Portugal. 
\and
        \label{aip}Leibniz-Institut f\"ur Astrophysik Potsdam (AIP), An der Sternwarte 16, D-14482 Potsdam, Germany.
        \and
\label{prague}Astronomical Institute, Academy of Sciences of the Czech Republic, Bo\v{c}n\'{i} II 1401/1a, CZ-141 00 Prague, Czech Republic.
\and
\label{brno}Department of Theoretical Physics and Astrophysics, Faculty of Science, Masaryk University, Kotl\'{a}\v{r}sk\'a 2, CZ-611\,37 Brno, Czech Republic
 \and
\label{mpia}Max-Planck-Institut f\"ur Astronomie, Heidelberg, Germany. 
\and
        \label{sydney}Sydney Institute for Astronomy, School of Physics A28, University of Sydney, NSW 2006, Australia.
        \and
        \label{aao}Australian Astronomical Observatory, PO BOX 296, Epping, NSW 1710, Australia.
\and
\label{porto}Centro de Astrof\'\i sica and Faculdade de Ciencias,
Universidade do Porto, Rua das Estrelas, 4150-762 Porto, Portugal.
     \thanks{Based on observations collected at the Centro Astron\'omico
      Hispano Alem\'an (CAHA) at Calar Alto, operated jointly by the Max-Planck
      Institut f\"ur Astronomie and the Instituto de Astrof\'{\i}sica de Andaluc\'{\i}a (CSIC).}
              }

   \date{Received ----- ; accepted ---- }

 
\abstract{ 

We present the largest and most homogeneous catalog of \ion{H}{ii} regions and
associations compiled so far. The catalog comprises more than 7000
ionized regions, extracted from 306 galaxies observed by the CALIFA survey.
We describe the procedures used to detect, select, and
analyse the spectroscopic properties of these ionized regions.
In the current study we focus on the characterization of the radial
gradient of the oxygen abundance in the ionized gas, based on the
study of the deprojected distribution of \ion{H}{ii} regions. We found
that all galaxies without clear evidence of an interaction present a
common gradient in the oxygen abundance, with a characteristic slope
of $\alpha_{O/H}=-$0.1 dex/$r_e$ between 0.3 and 2 disk effective
radii ($r_e$), and a scatter compatible with random fluctuations
around this value, when the gradient is normalized to the disk
effective radius. The slope is independent of morphology, incidence of bars, absolute magnitude or
mass. Only those galaxies with evidence of interactions and/or clear
merging systems present a significant shallower gradient, consistent
with previous results.  The majority of the 94 galaxies with
\ion{H}{ii} regions detected beyond 2 disk effective radii present a
flattening in the oxygen abundance. The flattening is statistically
significant. We cannot provide with a conclusive answer regarding the
origin of this flattening. However, our results indicate that its
origin is most probably related to the secular evolution of galaxies.
Finally, we find a drop/truncation of the oxygen abundance in the
inner regions for 26 of the galaxies. All of them are non-interacting,
mostly unbarred, Sb/Sbc galaxies. This feature is associated with a
central star-forming ring, which suggests that both features are
produced by radial gas flows induced by resonance processes.  Our
result suggest that galaxy disks grow inside-out, with metal
enrichment being driven by the local star-formation history, and with
a small variation galaxy-by-galaxy. At a certain galactocentric
distance, the oxygen abundance seems to be well correlated with the
stellar mass density and total stellar mass of the galaxies,
independently of other properties of the galaxies. Other processes,
like radial mixing and inflows/outflows, although they are not ruled
out, seem to have a limited effect on shaping of the radial
distribution of oxygen abundances.  }

\keywords{Galaxies: abundances --- Galaxies: fundamental parameters ---
  Galaxies: ISM --- Galaxies: stellar content --- Techniques: imaging
  spectroscopy --- techniques: spectroscopic -- stars: formation -- galaxies:
  ISM -- galaxies: stellar content}

\maketitle


\section{Introduction}
\label{intro}



The nebular emission arising from extragalactic objects has played an
important role in the new understanding of the Universe and its
constituents brought about by the remarkable flow of data over the last few
years, thanks to large surveys such as the 2dFGRS
\citep{1999MNRAS.308..459F}, SDSS \citep{York:2000p2677}, GEMS
\citep{Rix:2004p3993} or COSMOS \citep{2007ApJS..172....1S}, to name a
few.  Nebular emission lines have been, historically, the main tool at
our disposal for the direct measurement of the gas-phase abundance at
discrete spatial positions in low redshift galaxies. They trace the
young, massive star component in galaxies, illuminating and ionizing
cubic kiloparsec sized volumes of interstellar medium (ISM). Metals
are a fundamental parameter for cooling mechanisms in the
intergalactic and interstellar medium, star formation, stellar
physics, and planet formation. Measuring the chemical abundances in
individual galaxies and galactic substructures, over a wide range of
redshifts, is a crucial step to understanding the chemical evolution
and nucleosynthesis at different epochs, since the chemical abundance
pattern trace the evolution of past and current stellar generations.
This evolution is dictated by a complex set of parameters, including
the local initial gas composition, star formation history (SFH), gas
infall and outflows, radial transport and mixing of gas within disks,
stellar yields, and the initial mass function. The details of these
complex mechanisms are still not well established observationally, nor well developed theoretically, and hinder our understanding of
galaxy evolution from the early Universe to present day. 

Previous spectroscopic studies have unveiled some aspects of the complex
processes at play between the chemical abundances of galaxies and
their physical properties. Although these studies have been successful
in determining important relationships, scaling laws and systematic
patterns (e.g.\ luminosity-metallicity, mass-metallicity, and surface
brightness vs. metallicity relations
\citealt{leque79,Skillman:1989p1592,VilaCostas:1992p322,zaritsky94,tremonti04};
effective yield vs. luminosity and circular velocity relations
\citealt{Garnett:2002p339}; abundance gradients and the effective
radius of disks \citealt{diaz89}; systematic differences in
the gas-phase abundance gradients between normal and barred spirals
\citealt{zaritsky94,Martin:1994p1602}; characteristic
vs. integrated abundances \citealt{moustakas06}; etc.), they
have been limited by statistics, either in the number of observed \ion{H}{ii}
regions or in the coverage of these regions across the galaxy
surface.

The advent of Multi-Object Spectrometers and Integral Field
Spectroscopy (IFS) instruments with large fields of view now offers us
the opportunity to undertake a new generation of emission-line
surveys, based on samples of hundreds of \ion{H}{ii} regions and
with full two-dimensional (2D) coverage of the disks of nearby spiral
galaxies. In the last few years we started a major observational program to
understand the statistical properties of \ion{H}{ii} regions, and to
unveil the nature of the reported physical relations, using IFS. This
program was initiated with the PINGS survey \citep{rosales-ortega10},
which acquired IFS mosaic data for a number of medium sized nearby galaxies. 
We then continued the acquisition of IFS data for a larger sample of visually
classified face-on spiral galaxies \citep{marmol-queralto11}, as part of the
feasibility studies for the CALIFA survey \citep{sanchez12a}. 

In \cite{sanchez12b} we presented a new method to detect, segregate
and extract the main spectroscopic properties of \ion{H}{ii} regions
from IFS data
(\textsc{HIIexplorer}\footnote{\url{http://www.caha.es/sanchez/HII_explorer/}}).
Using this tool, we have built the largest and homogenous catalog of
\ion{H}{ii} regions for the nearby Universe. This catalog has allowed
us to establish a new scaling relation between the local stellar mass
density and oxygen abundance, the so-called \Sz\ relation
\citep{rosales12}, and to explore the galactocentric gradient of the
oxygen abundance \citep{sanchez12b}.  We confirmed that up to $\sim$2
disk effective radii there is a negative gradient of the oxygen
abundance in all the analyzed spiral galaxies. This result is in
agreement with models based on the standard inside-out scenario of
disk formation, which predict a relatively quick self enrichment with
oxygen abundances and an almost universal negative metallicity
gradient once this is normalized to the galaxy optical size
\citep{bois99,bois00}.  Indeed, the measured gradients present a very
similar slope for all the galaxies ($\sim -0.12$ dex/$r_e$), when the
radial distances are measured in units of the disk effective radii. We
found no difference in the slope for galaxies of different
morphological types: early/late spirals, barred/non-barred,
grand-design/flocculent.

Beyond $\sim$2 disk effective radii our data show evidence of a
flattening in the abundance, consistent with several
other spectroscopic explorations, based mostly on single objects
\citep[e.g.][]{bresolin09,yoac10,rosales11,mari11,bresolin12}. The same
pattern in the abundance has been described in the case of the
extended UV disks discovered by GALEX \citep{gil05,thil07}, which show
oxygen abundances that are rarely below one-tenth of the solar value.
Additional results, based on the metallicity gradient of the outer
disk of NGC 300 from single-star CMD analysis \citep{vlaj09} support the
presence of a flatter gradient towards the outer disks of spiral galaxies. 
In the case of the Milky Way (MW), studies using open clusters 
\citep[e.g.][]{2008A&A...480...79B,2009A&A...494...95M,2012AJ....144...95Y,pedi09},
Cepheids \citep[e.g.][]{2002A&A...392..491A,2003A&A...401..939L,2004A&A...413..159A,2008A&A...490..613L},
\hh regions \citep[e.g.][]{1996MNRAS.280..720V,este13},
 PNe \citep[e.g.][]{2004A&A...423..199C}, and a combination of different tracers \citep[e.g.][]{maciel09} also report a flattening of the
gradient in the outskirts of the Milky Way, somewhere between 10 and 14 kpc
\footnote{Although there are recent studies which do not favour a
flattening of the MW gradient in the outer disk based on Cepheids,
\citep[e.g.][and references therein]{2013A&A...558A..31L}.}. 
Despite all these results, the outermost parts of the disk have not been
explored properly, either due to the limited number of objects considered in the
previous studies, or due to the limited spatial coverage \citep[e.g.,][]{sanchez12b}.

\begin{figure}
\centering
\includegraphics[width=6.5cm,clip,trim=0 0 0 0,angle=270]{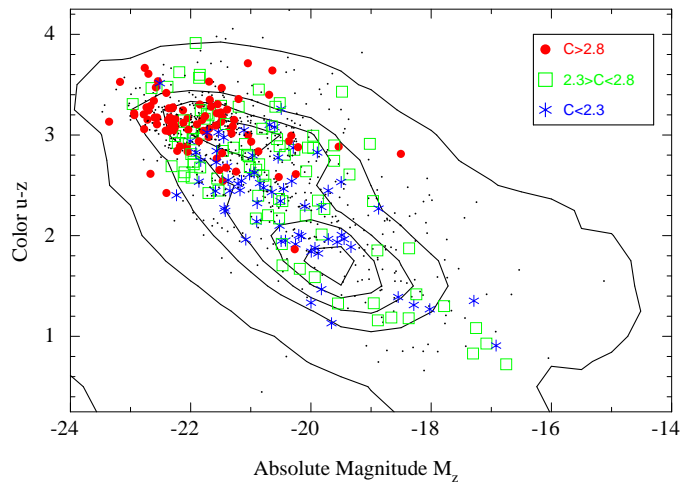}
\caption{\label{fig:CM} Distribution of the currently observed CALIFA galaxies
  in the $u-z$ vs.\ $M_z$ color-magnitude diagram. Different colors and
  symbols represent a classification into spheroid- and disk-dominated
  galaxies as well as intermediate cases, as suggested by the concentration
  index $C$ (see Sec. \ref{sample} for definition). For comparison, the
  contours delineate the number density distribution of galaxies in the
  SDSS-NYU catalogue \citep[e.g.][]{blanton05}. The small dots indicate the
  remaining objects in the CALIFA mother sample, not observed so far. }
\end{figure}

The search for an explanation of the existence of radial gradients of
abundances (and the G-dwarf metallicity distribution in MW) was the
reason for the early development of chemical evolution models as well
as the classical closed box model (CBM). The pure CBM, which relates
the metallicity or abundance of a region to its fraction of gas,
independently of the star formation or evolutionary history, was
unable to explain the radial abundance gradient observed in our Galaxy
and in other spirals. Therefore infall or outflows of gas in the MW
were considered necessary to fit the data. In fact, as explained by
\cite{goet92}, there are only 4 possible ways to create a radial
abundance gradient: 1) A radial variation of the initial mass function
(IMF); 2) A variation of the stellar yields with galactocentric
radius; 3) A star formation rate (SFR) changing with the radius; 4) A
gas infall rate variable with radius. The first possibility is not
usually considered as probable, and the second one is already included
in modern models, that adopt metallicity dependent stellar yields. Thus,
from the seminal works of \cite{lace85}, \cite{guest82} and
\cite{clay87}, most of numerical chemical evolution models
\citep[e.g.][]{diaz84,matte89,ferri92,cari94,prant95,molla96,chiap97,boiss99}
explain the existence of the radial gradient of abundances by the
combined effects of a star formation rate and an infall of gas, both
varying with galactocentric radius of galaxies. In most recent
times chemical evolution has been included in modern cosmological
simulation codes, which already obtain spiral disks as observed,
finding radial gradients of abundances which reproduce the data
\citep{pilki92}. It has been demonstrated \cite{gibson13}, that the
existence and evolution of these radial gradients is, as expected,
very dependent on the star formation and infall prescriptions included
in the simulations.

To characterize the properties of the ISM in the Local Universe and
their relations with the evolution of galaxies, we applied the previously
described procedure to the IFS data provided by the CALIFA survey
\citep{sanchez12a}\footnote{http://califa.caha.es/}. CALIFA is an
ongoing exploration of the spatially resolved spectroscopic properties
of galaxies in the Local Universe ($z<$0.03) using wide-field IFS to
cover the full optical extent (up to $\sim$3--4 r$_e$) of $\sim$600
galaxies of any morphological type, distributed across the entire
color-magnitude diagram (Walcher et al., in prep.), and sampling the
wavelength range 3650-7500 \AA. So far, the survey has completed
$\sim$1/2 of its observations, with 306 galaxies observed (May 2013),
and the first data release, comprising 100 galaxies, was 
delivered in November 2012 \cite{husemann13}.

In \cite{sanchez13} we presented the first results based on the
catalog of \HII\ regions extracted from these galaxies. We studied the
dependence of the \mz\ relation with the star-formation rate, finding
no secondary relation different from the one induced by the well known
relation between the star formation and the mass. We confirm the local
\Sz\ relation unveiled by \cite{rosales11}, with a larger statistical
sample of \HII regions.

In the current study we will use the updated CALIFA catalog of \HII
regions to study the radial oxygen abundance gradient up to 3-4
disk effective radii, well beyond the proposed break/flattening. The
layout of this article is as follows: in Sec.\ \ref{sample} we summarize
the main properties of the sample and data used in this study; in
Sec.\ \ref{ana} we describe the analysis required to detect the
individual clumpy ionized regions and aggregations, and to extract
their spectroscopic properties, in particular the emission line ratios
required to determine the abundance; 
the criteria to select the \ion{H}{ii} region are explained in \ref{HII_select};
the derivation of the abundance gradient for each galaxy is 
described in Sec.\ \ref{OH}; in Sec.\ \ref{morph} we explore the dependence of
the slope of these gradients with different morphological and structural 
properties of the galaxies; in Sec. \ref{common} we describe the properties
of the common gradient of the oxygen abundance for all disk galaxies up to
$\sim$2 $r_{e}$, and the presence of a flattening beyond this radius; the drop
of the abundance for some particular galaxies is shown in
Sec.\ \ref{drop}. Finally, the main conclusions of this study are discussed in
Sec.\ \ref{disc}.


\section{Sample of galaxies and dataset}\label{sample}

The galaxies were selected from the CALIFA observed sample. Since
CALIFA is an ongoing survey, whose observations are scheduled on a
monthly basis (i.e., dark nights), the list of objects increases
regularly. The current results are based on the 306 galaxies observed
before May 2013, i.e., half of the foreseen 600 galaxies to be observed
at the end of the survey. Figure \ref{fig:CM} shows the distribution
of the current sample along the color-magnitude diagram, indicating
with different symbols galaxies of different concentration index $C$
(defined to be the ratio $C = R_{90}/R_{50}$, where $R_{90}$ and $R_{50}$ are
the radii enclosing 90\% and 50\% of the Petrosian r-band luminosity of the
galaxy; i.e., a proxy of the morphological type). The current sample covers all the
color-magnitude diagram, up to $M_{z}<-17$ mag, with at least
three targets per bin of $\sim$1 magnitude and color ($\sim$10
galaxies on average) including galaxies of any morphological type. The
CALIFA mother sample becomes incomplete below $M_r>-19$ mag, which
corresponds to a stellar mass of $\sim$10$^{9.5}$ M$_\odot$ for a
Chabrier IMF. Therefore, it does not sample properly the low-mass
and/or dwarf galaxies. Above this luminosity, the sample is
representative of the total population at the selected redshift range
(0.005$<z<$0.03), and in principle it should be representative
of galaxies at the Local Universe.


\begin{figure*}
\centering
\includegraphics[width=8cm,clip,trim=0 20 60 0]{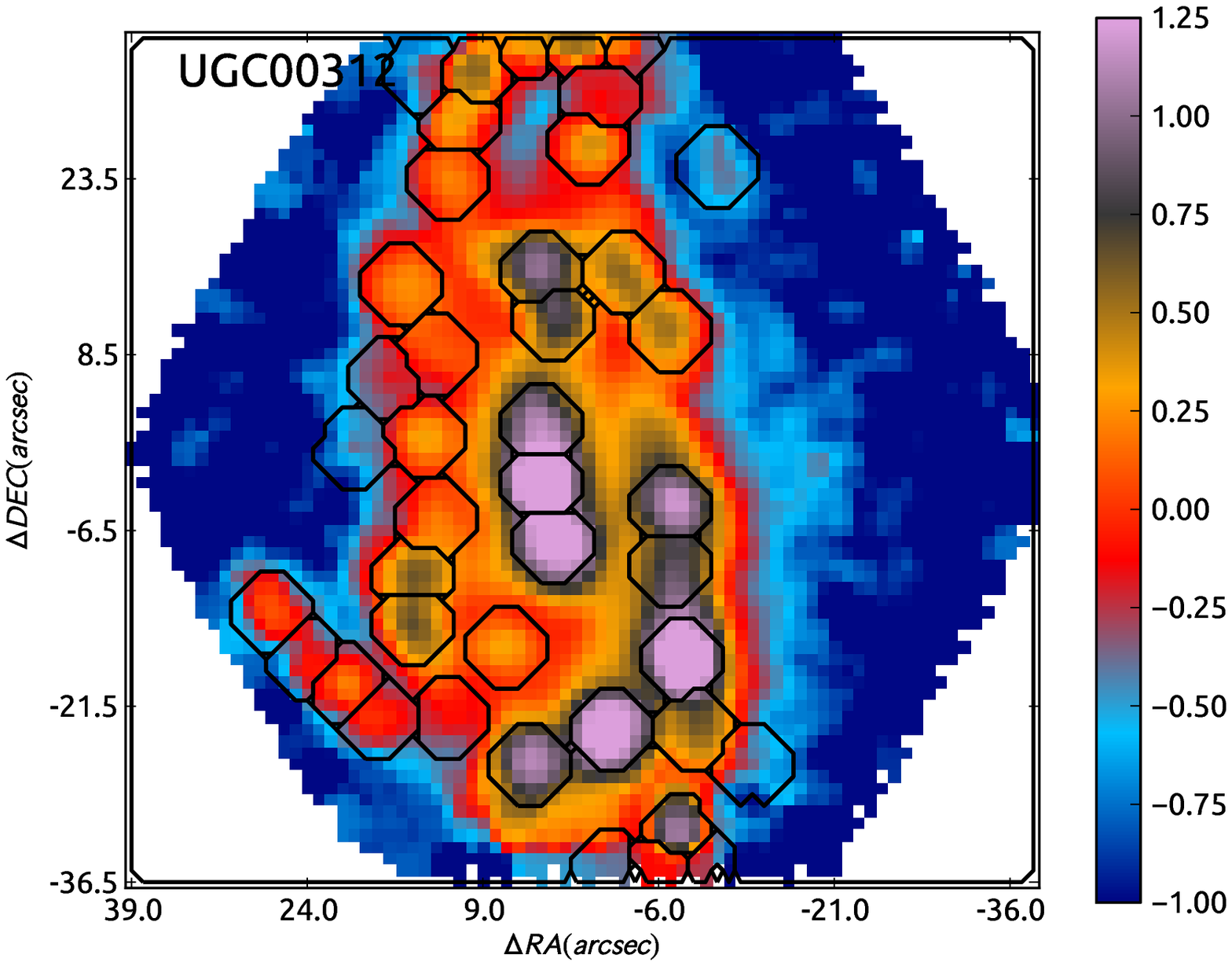}
\includegraphics[width=8cm,clip,trim=0 20 60 0]{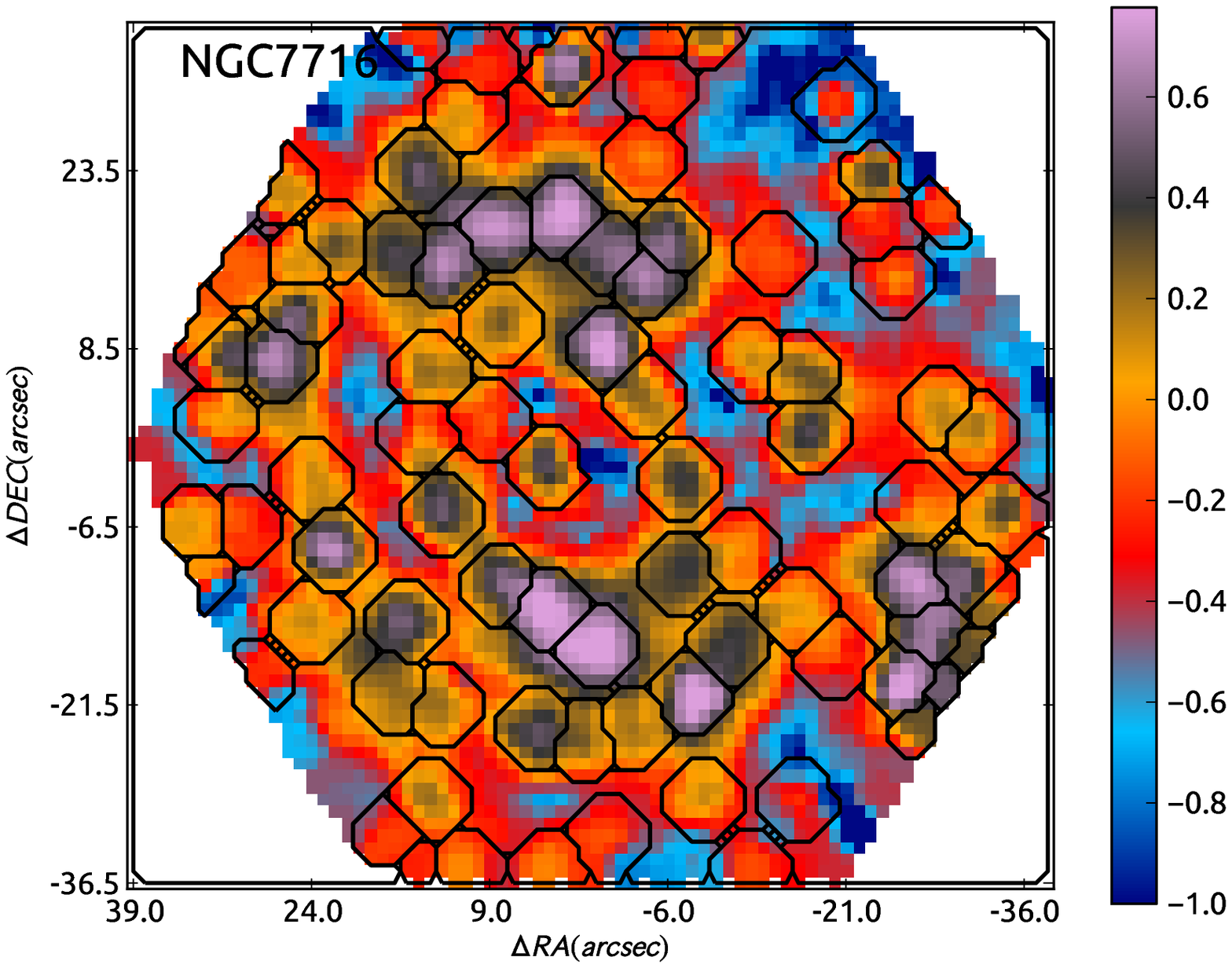}
\caption{\label{fig:HII} IFS-based H$\alpha$ maps, in units of ($\log_{10}$)
  \FunitsA, derived for two representative galaxies of the  sample,
  together with the detected \ion{H}{ii} regions shown as black segmented
  contours.}
\end{figure*}

The details of the survey, sample, observational strategy, and
reduction are explained in \cite{sanchez12a}. All galaxies were
observed using PMAS \citep{roth05} in the PPAK configuration
\citep{kelz06}, covering a hexagonal field-of-view (FoV) of
74$\arcsec$$\times$64$\arcsec$, sufficient to map the full optical
extent of the galaxies up to 2-3 disk effective radii. This is
possible because of the diameter selection of the sample (Walcher et
al., in prep.). The observing strategy guarantees a complete coverage
of the FoV, with a final spatial resolution of FWHM$\sim$3$\arcsec$,
corresponding to $\sim$1 kpc at the average redshift of the
survey. The sampled wavelength range and spectroscopic resolution
(3745-7500\AA, $\lambda/\Delta\lambda\sim$850, for the low-resolution
setup) are more than sufficient to explore the most prominent ionized
gas emission lines, from [OII]$\lambda$3727 to [SII]$\lambda$6731, on
one hand, and to deblend and subtract the underlying stellar
population, on the other hand
\citep[e.g.][]{sanchez12a,kehrig12,cid-fernandes13}. The dataset was
reduced using version 1.3c of the CALIFA pipeline, whose modifications
with respect to the one presented in \cite{sanchez12a} are described
in detail in \cite{husemann13}. In summary, the data fulfil the
predicted quality-control requirements, with a spectrophotometric
accuracy better than 15\%\ everywhere within the wavelength range,
both absolute and relative, and a depth that allows us to detect
emission lines in individual \HII\ regions as weak as
$\sim$10$^{-17}$\flux, with a signal-to-noise ratio of
S/N$\sim$3-5. For the emission lines considered in the current study
the S/N is well above this limit, and the measurement errors are
neglectable in most of the cases. In any case, they have been
propagated and included in the final error budget.

The final product of the data-reduction is a regular-grid datacube,
with $x$ and $y$ coordinates indicating the right-ascension and
declination of the target and $z$ being a common step in
wavelength. The CALIFA pipeline also provides  the propagated error
cube, a proper mask cube of bad pixels, and a prescription of how to
handle the errors when performing spatial binning (due to covariance
between adjacent pixels after image reconstruction). These datacubes,
together with the ancillary data described in Walcher et al. (in
preparation), are the basic starting point of our analysis.

The observing strategy of the CALIFA survey guarantees that the main
properties of the observed sample are compatible with those of the
mother sample, in terms of luminosities, sizes, morphologies and
colors \citep{sanchez12a,husemann13}. Particular care is taken 
to not introduce any potential observational bias, as the targets are
selected in a pseudo-random way based only on the visibility from the
observatory on a monthly basis (i.e., dark-time). In Walcher et
al. (in prep.), we will describe the main properties of the CALIFA
mother sample. In summary we can claim that with our selection
criteria our sample does not under-represent any kind of galaxies in the
Local Universe in any observable within our 95\% completeness range
($-$23$<$M$_{r,SDSS}$$<-$19 mag). Obviously our results are restricted
to this particular range, and therefore cannot be applied to either
dwarf or giant elliptical galaxies, which are under-represented or absent
in our sample.

\section{Analysis}\label{ana}

The main goal of this study is to characterize the
abundance gradient in galaxies, and determine if there are common patterns
and/or differences depending on their individual properties. Ionized gas
abundances have been well calibrated on the basis of strong-line indicators
for ionized regions associated with star formation processes, i.e., the
classical \ion{H}{ii} regions. In this section we describe how we have
selected those regions, extract and analyze their individual spectra,
derive the corresponding oxygen abundance and, finally, analyze their radial
gradient.

\subsection{Detection of ionized regions}\label{HII_detect}

The segregation of \ion{H}{ii} regions and the extraction of the
corresponding spectra is performed using a semi-automatic procedure
named {\sc HIIexplorer}\footnote{\url{http://www.caha.es/sanchez/HII_explorer/}}. The
procedure is based on some basic assumptions: (a) \ion{H}{ii} regions
are peaky/isolated structures with a strong ionized gas emission, that is
significantly above the stellar continuum emission and the average ionized gas
emission across the galaxy. This is particularly true for 
H$\alpha$; (b) \ion{H}{ii} regions have a typical physical
size of about a hundred or a few hundred parsecs
\citep[e.g.][]{rosa97,lopez2011,oey03}, which corresponds to a typical
projected size of a few arcsec at the  distance  of the galaxies.

These basic assumptions are based on the fact that most of the H$\alpha$
luminosity observed in spiral and irregular galaxies is a direct
tracer of the ionization of the inter-stellar medium (ISM) by the
ultraviolet (UV) radiation produced by young high-mass OB
stars. Since only high-mass, short-lived, stars contribute
significantly to the integrated ionizing flux, this luminosity is a
direct tracer of the current star formation rate (SFR), independent of the
previous star formation history. Therefore, clumpy structures detected in
the H$\alpha$ intensity maps are most probably
associated with classical \ion{H}{ii} regions (i.e., those regions for which the oxygen
abundances have been calibrated).

The details of {\sc HIIexplorer} are given in \cite{sanchez12b} and
\cite{rosales12}. We present here the basic steps included in the
overall process: (i) First we create a narrow-band image of 120$\AA$
width, centered on the wavelength of H$\alpha$ shifted at the redshift
of each target. The image was created by co-adding the flux within the
described spectral window for each spaxel of the velocity-field
corrected datacube. Then, the image is properly corrected for the
underlying adjacent continuum. (ii) This image is used as an input for
the automatic \ion{H}{ii} region detection algorithm included in
\textsc{HIIexplorer}. In this particular case, the algorithm detects
iteratively the peak intensity emission above a threshold of
4$\times$\fuden, and then assigns all
the adjacent pixels up to a distance of 3.5$\arcsec$, with a flux
within a 10\% of the peak intensity ($I_{pixel}>0.9\times I_{peak}$),
and above a limiting flux intensity of 1$\times$\fuden\ into the corresponding area. Once the first
region is detected and segregated, the corresponding area is masked
from the input image, and the procedure is repeated until there are no
additional regions to be selected. The remaining pixels are assigned
to a residual region which is assumed to be dominated by diffuse
emission. The result is a segmentation map that segregates each
detected cumpy ionized structure. Finally, (iii) the integrated
spectra corresponding to each segmented region is extracted from the
original datacube, and the corresponding position table of the
detected \ion{H}{ii} is provided. If the object was has been observed in
both  the low-resolution and high-resolution modes
\citep{sanchez12a}, both corresponding spectra were extracted.

\begin{figure*}
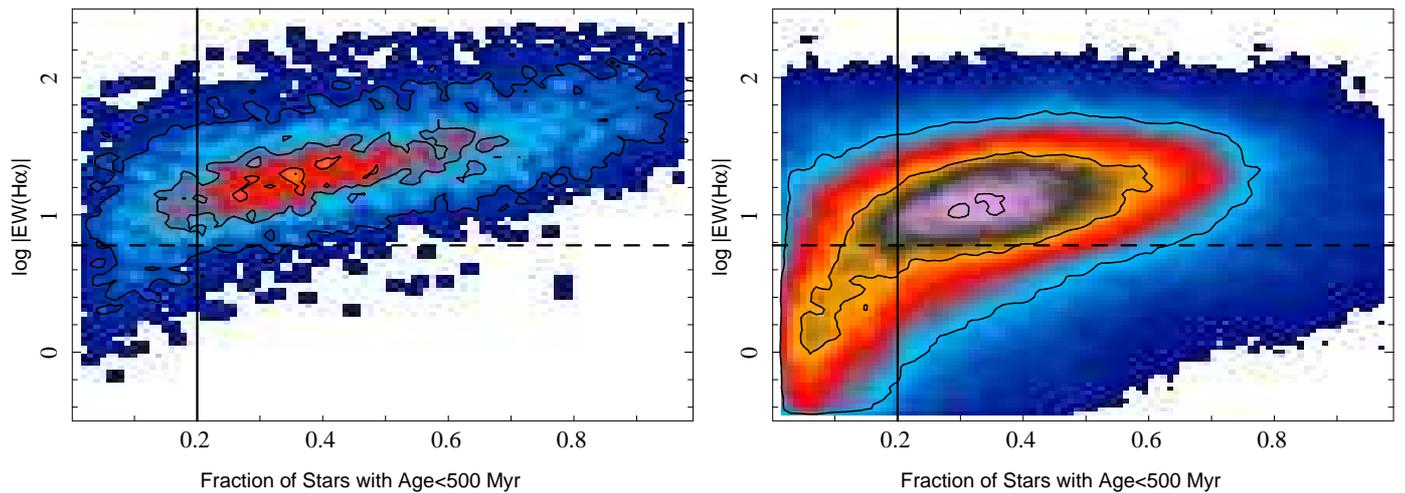

\centering
\includegraphics[width=6.7cm,angle=270]{figs/frac_young_log_EW.ps}
\includegraphics[width=6.7cm,angle=270]{figs/f_y_EW_CS.ps}
\caption{\label{fig:EW} {\it Left-panel:} Absolute value of the equivalent
  width of H$\alpha$, on a logarithmic scale, plotted against the fraction of
  young stars in the underlying stellar population (according to the SSP
  modelling by {\tt FIT3D}) for the clumpy ionized regions selected with {\sc
    HIIexplorer}. {\it Right-panel:} Similar distribution of the emission
  equivalent width of H$\alpha$, on a logarithmic scale, along the luminosity fraction of
  young stars in the underlying stellar population for the $\sim$500,000
  individual spaxels with H$\alpha$ emission detected in the 306 analysed
  datacubes. In both panels, the dashed horizontal line shows the demarcation
  limit of the EW of H$\alpha$ proposed by \cite{cid-fernandes10} to
  distinguish between strong (star formation and/or AGNs) and weak (e.g.,
  post-AGBs) ionization sources. The vertical solid line shows the currently
  adopted selection criteria for the \ion{H}{ii} regions.}
\end{figure*}

Figure \ref{fig:HII} illustrates the process, showing the H$\alpha$
intensity maps and the corresponding segmentations for two objects:
(1) UGC00312, an intermediate-to-high inclined ($\sim$70 degrees), not
very massive ($\sim$0.7$\times$10$^{10}$ M$_\odot$) and almost bulge-less spiral
galaxy, and (2) NGC7716, a low inclination ($<$40 degrees), massive
($\sim$2$\times$10$^{10}$ M$\odot$) spiral, with a clear
bulge. 
These two galaxies illustrate why for highly inclined galaxies we
cover up to 4-5 disk effective radii (mostly along the semi-minor axis), while for mostly face-on ones we
cover just half of this size, although both galaxies were diameter selected \citep{sanchez12a}. The galaxies analyzed in
\cite{sanchez12b}, are more similar to the second type, and therefore
the region beyond $\sim$2 effective radii was mostly unexplored.

A total of 7016 individual clumpy ionized regions are detected in a
total of 227 galaxies from the sample, i.e., $\sim$30 \ion{H}{ii}
regions per galaxy. This does not mean that on averarage there is no
ionized gas in the remaining 79 galaxies. Recent results indicate that
it is possible to detect low-intensity (and in most cases
low-ionization) gas in all the analyzed CALIFA galaxies
\citep[][]{kehrig12,papa13,2013arXiv1308.4271S}.  As discussed in
\citet{papa13}, and in line with a substantial body of previous works
\citep[][among
  others]{2010MNRAS.402.2187S,2010A&A...519A..40A,2012ApJ...747...61Y,kehrig12},
various lines of evidence suggest that photoionization by post-AGB
stars appears to be the main driver of extended nebular emission in
these systems, with non-thermal sources being potentially important
only in their nuclei. The observational evidence behind this
conclusion is that the nebular emission is not confined only to the
nuclear regions but is extended out to r $\sim$2--4 r$_{50}$, {
  i.e.\ it is co-spatial with the post-AGB stellar background. In most
  of these cases EW(H$\alpha$) typically is $\sim$\,1\,\AA.  In other
  cases the ionized gas does not present clear clumpy structures,
  required to associate them with starforming/\ion{H}{ii}
  regions. This is the case of the shock ionized regions detected in
  the MICE galaxies (Wild et al., submitted). Since most of this
  ionization is not associated with young massive stars, and therefore
  the associated abundances are not well calibrated, it is not
  relevant for the present analysis. }


\subsection{Measurement of the emission lines}\label{HII_elm}

To extract the nebular physical information of each individual
\ion{H}{ii} region, the underlying stellar continuum must be decoupled
from the emission lines for each of the analyzed spectra. Several
different tools have been developed to model the underlying stellar
population, effectively decoupling it from the emission lines
\citep[e.g.,][]{cappellari04,cid-fernandes05,ocvrik2006,sarzi2006,sanchez06b,koleva2009,macarthur2009,walcher11}.
Most of these tools are based on the same principles, i.e., they
assume that the stellar emission is the result of the combination of
different (or a single) simple stellar populations (SSP), and/or the
result of a particular star-formation history, whose corresponding
emission-line spectrum is redshifted due to a certain systemic
velocity, broadened and smoothed by the effect of a certain velocity
dispersion and attenuated by a certain dust content.

We performed a simple modeling of the continuum emission using {\tt
  FIT3D}\footnote{\url{http://www.caha.es/sanchez/FIT3D/}}, a fitting
package described in \cite{sanchez06b} and \cite{sanchez11}.  A simple
SSP template grid with 12 individual populations was adopted.  It
comprises four stellar ages (0.09, 0.45, 1.00 and 17.78 Gyr), two
young and two old ones, and three metallicities (0.0004, 0.019 and
0.03), sub-solar, solar and super-solar. The models were extracted
from the SSP template library provided by the MILES project
\citep{vazdekis10,falc11}. The use of different stellar ages and
metallicities or a larger set of templates does not affect
qualitatively the derived quantities that describe the stellar
populations. Evenmore, it does not affect quantitatively the
estimations of the properties of the emission lines.


The analysis of the underlying stellar population is not as detailed as the
one presented by \cite{cid-fernandes13}, and it is
not useful to reconstruct the star formation history.
However,
since the spatial binning required to define these regions is based on
the H$\alpha$ intensity, in many cases the extracted spectra of the
underlying stellar continuum do not reach the required
signal-to-noise to perform a more detailed analysis. We prefer to
restrict our stellar fitting to a reduced template library with few
stellar populations, and derive simple conclusions, such as the fraction of young or old
stars. Therefore, we will not pay too much attention to the actual decomposition in
different populations.

Throughout, we adopted the \cite{cardelli89} law for the stellar dust
attenuation with an specific attenuation of $R_{\rm V}=3.1$, assuming
a simple screen distribution. The use of different laws, like the one
proposed by \cite{calz01}, does not produce significant differences in
the modelling of the underlying stellar population in the 
wavelength range considered. A different amount of extinction, parametrized by
the extinction in the V-band ($A_V$), was considered for each stellar
population. We consider that this is more realistic than to assume the
same attenuation for all the stellar populations, since the
distribution of the dust grains is not homogeneous, and it affects  the old and young stellar populations in a
different way.

Individual emission line fluxes were measured using {\tt FIT3D} in the
{\it stellar-population subtracted} spectra performing a multi-component
fitting using a single Gaussian function. When more than one emission
line was fitted simultaneously (e.g., for doublets and triplets, like the
[\ion{N}{ii}] lines), the systemic velocity and velocity dispersion were
forced to be equal, in order to decrease the number of free parameters and
increase the accuracy of the deblending process. The ratio between the two
\nii\ lines included in the spectral range were fixed to the theoretical value
\citep{Osterbrock:2006p2331}. By adopting this procedure it is possible to
accurately deblend the different emission lines. A similar procedure was
applied to the rest of the lines which were fitted simultaneously
(e.g. H$\beta$ and \oiii). The measured lines include all lines employed in
the determination of metallicity using strong-line methods, i.e \ha, \hb, 
\oii\ \lam3727, \oiii\ \lam4959, \oiii\ \lam5007, \nii\ \lam6548, \nii\ \lam6583,
\sii\ \lam6717 and \sii\ \lam6731.
Additionally, for those \hh regions with high signal-to-noise we were able to
detect and measure intrinsically fainter lines such as
[Ne\,{\footnotesize III}] \lam3869, H$\epsilon$ \lam3970, H$\delta$ \lam4101, \hg\ \lam4340,
He\,{\footnotesize I} \lam5876, [O\,{\footnotesize I}] \lam6300, and
He\,{\footnotesize I} \lam6678, although they have not been considered for
the present study.
{\tt FIT3D} provides the intensity, equivalent width, systemic velocity and
velocity dispersion for each emission line. The statistical uncertainties in
the measurements were calculated by propagating the error associated with the
multi-component fitting and considering the signal-to-noise of the spectral
region. Note that by subtracting a stellar
continuum model derived with a set of SSP templates, we are already correcting
for the effect of underlying stellar absorption, which is particularly
important in Balmer lines (such as H$\beta$). We performed a series of sanity
tests based on the \ha/\hb\ ratio to ensure that no overcorrection was done on
the absorption stellar features.

Note that {\tt FIT3D} fits  the underlying stellar population and the
emission lines together. Therefore, in addition to the parameters derived for the
emission lines, the fitting algorithm provides us with a set of
parameters describing the physical components of the stellar
populations. In particular it provides the fraction of light 
 that contributes to the continuum at
5000\AA\ corresponding to an old ($>$500 Myr, $f_o$) or young ($<$500
Myr, $f_y$) stellar population (which we consider a reliable parameter for
our current stellar analysis).

\begin{figure*}
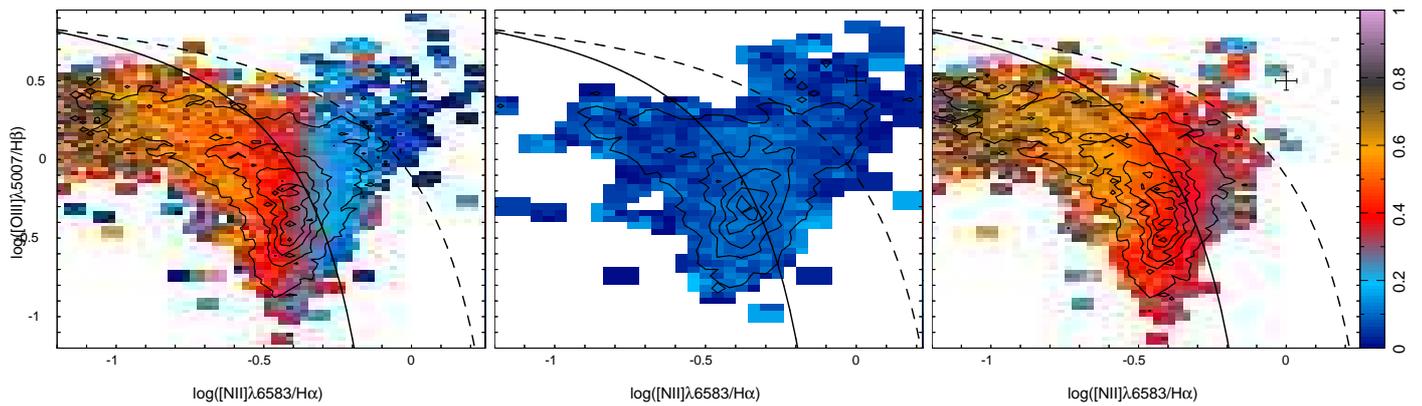

\centering
\includegraphics[width=5.6cm,angle=270,clip,trim=0 7 0 80]{figs/diag_O3_N2_Frac.ps}~
\includegraphics[width=5.6cm,angle=270,clip,trim=0 72 0 80]{figs/diag_O3_N2_Frac_Old.ps}~
\includegraphics[width=5.6cm,angle=270,clip,trim=0 72 0 0]{figs/diag_O3_N2_Frac_Young.ps}
\caption{\label{fig:BPT} {\it Left panel:}
  [\ion{O}{iii}]~$\lambda$5007/H$\beta$
  vs. [\ion{N}{ii}]~$\lambda$6583/H$\alpha$ diagnostic diagram for the
  $\sim$7000 ionized regions described in the text. The contours show the
  density distribution of these regions with the diagram plane, with the
  outermost contour enclosing 95\% of the regions, and each consecutive one
  enclosing 20\% less regions. The color indicate the fraction of young
  stellar population in the underlying continuum. {\it Central panel:} Same
  diagnostic diagram, restricted to those ionized regions with less than a
  20\% of young stellar population ($\sim$1800 regions). {\it Right panel:}
  Same diagnostic diagram, restricted to those ionized regions with more than
  a 20\% of young stellar population ($\sim$5800 regions). In all the panels,
  the solid- and dashed-line represent, respectively, the \cite{kauffmann03}
  and \cite{kewley01} demarcation curves. They are usually invoked to
  distinguish between classical star-forming objects (below the solid-line),
  and AGN powered sources (above the dashed-line). Regions between both lines
  are considered intermediate ones.}
\end{figure*}

\subsection{Selection of the \ion{H}{ii} regions}\label{HII_select}

Classical \ion{H}{ii} regions are gas clouds ionized by short-lived
hot OB stars, associated with on-going star formation. They are frequently
selected on the basis of demarcation lines defined in the so-called
diagnostic diagrams
\citep[e.g.][]{baldwin81,veilleux87}, which compare different line ratios, such as
[OIII]/H$\beta$ vs [NII]/H$\alpha$, [OIII]/H$\beta$ vs
[OII]/H$\alpha$, [NII]/H$\alpha$ vs. [SII]/H$\alpha$ and/or
[NII]/H$\alpha$ vs. [SII]/$H\alpha$. In most  cases
these ratios discriminate well between strong ionization sources, like
classical \ion{H}{ii} regions and powerful AGNs
\citep[e.g.][]{baldwin81}. However, they are less accurate in
distinguishing between low-ionization sources, like weak AGNs, shocks
and/or post-AGBs stars \citep[e.g.][]{cid11,kehrig12}. Alternative
methods, based on a combination of the classical line ratios with
additional information regarding the underlying stellar population
have been proposed. For example, \cite{cid11} proposed the use of the
EW(H$\alpha$), to distinguish between retired (non starforming) galaxies, weak
AGNs and star-forming galaxies.

The most common diagnostic diagram in the literature for the optical regime is
the one which makes use of easily-observable strong lines that are less
affected by dust attenuation, i.e.,
[\ion{O}{iii}]/H$\beta$ vs. [\ion{N}{ii}]/H$\alpha$ \citep{baldwin81}.
We will refer hereafter to this diagnostic diagram as the BPT diagram. Different demarcation lines have been proposed for
this diagram. The most popular ones are the \cite{kauffmann03} and
\cite{kewley01} curves. They are usually invoked to distinguish
between star-forming regions \citep[below the ][ curve]{kauffmann03},
and AGNs \citep[above the ][ curve]{kewley01} . The location between
both curves is normally assigned to a mixture of different sources of
ionization. Additional demarcation lines have been proposed for the
region above the \cite{kewley01} curve to segregate between Seyfert
and LINERs \citep[e.g.,][]{kewley06}.

Despite of the benefits of this {\it clean} segregation for
classification purposes, it may introduce biases when applied in order to select
\ion{H}{ii} regions. The \cite{kewley01} curve was derived on the
basis of photoionization models. It corresponds to the maximum
envelope in the considered plane for ionization produced by hot
stars. Therefore, to the extent that these models are realistic enough, any
combination of line ratios below this curve can be produced entirely
by OB star photoionization. Finally, it defines all the area above it
as {\it un-reachable} by ionization associated with star-formation.
The \cite{kauffmann03} curve has a completely different origin. It is
an empirical envelope defined to segregate between star-forming
galaxies and the {\it so-called} AGN branch in the BPT diagram based
on the analysis of the emission lines for the SDSS galaxies. It
describes well the envelope of classical \ion{H}{ii} regions found in
the disks of spiral galaxies. However, it is known that certain
\ion{H}{ii} regions can be found above this demarcation line, as we
will show below.


\cite{kennicutt89} first recognized that \ion{H}{ii} regions in the
center of galaxies distinguish themselves spectroscopically from those in the disk
 by their stronger low-ionization forbidden emission. The nature
of this difference was not clear. It may be due to contamination by an
extra source of ionization, like diffuse emission or the presence of
an AGN. However, other stellar processes, such as nitrogen enhancement
due to a natural aging process of \ion{H}{ii} regions and the
surrounding ISM can produce the same effect. These early results were
confirmed by \cite{ho97}, who  demonstrated that inner star-forming regions may
populate the right branch of the BPT diagram, at a location above the
demarcation line defined later by \cite{kauffmann03}. However, we
have found that these \ion{H}{ii} regions are not restricted to the
central regions, and can be found at any galactocentric distance,
even at more than 2 $r_e$ (Sec. \ref{common}), which excludes the
contamination by a central source of ionization. The nature of these
\ion{H}{ii} regions will be addressed in detail elsewhere. For the
purpose of the current study it is important to define  a selection
criterion that does not exclude them.


Therefore, selecting \ion{H}{ii} regions based on the
\cite{kauffmann03} curve may bias our sample towards classical disk
regions, excluding an interesting population of these objects. On the
other hand it does not guarantee the exclusion of other sources of
non-stellar ionization that can populate this area, like shocks
\cite[e.g.][]{allen04,leve10}, post-AGB stars \citep[e.g.][]{kehrig12}
and dusty AGNs \citep[e.g.][]{grov04}. Following
\cite{cid-fernandes10} and \cite{cid11}, we consider that an
alternative method to distinguish between different sources of
ionization is to compare the properties of the ionized gas with that
of the underlying stellar population. 

We adopted a different selection criterion, using the fraction of young
stars ($f_y$) provided by the multi-SSP analysis of the underlying
stellar population, as a proxy for the star-formation activity. 
For star-forming regions this parameter provides  similar information to the
EW(H$\alpha$). Figure \ref{fig:EW},
left-panel, shows the distribution of EW(H$\alpha$) against the
fraction of young stars for the $\sim$7000 clumpy ionized regions
selected by {\sc HIIexplorer}. For those regions with
EW(H$\alpha$)$>$6 \AA, and/or with a fraction of young stars larger
than 20\%, both parameters present a strong log-linear correlation
($r_{corr}=$0.95). Fig \ref{fig:EW}, right-panel, shows the same
distribution for the $\sim$500,000 spaxels with detected H$\alpha$
emission. This distribution presents the same trend described before,
but with an evident tail towards lower EW(H$\alpha$) values and a
lower fraction of young stars.

The threshold imposed by {\sc HIIexplorer} in the surface brightness
of H$\alpha$ and the requirement that the ionization is clumpy efficiently removes
 most of the ionization corresponding to weak-emission
lines described. This is mostly diffuse emission, that peaks in the
described diagram at EW(H$\alpha$)$\sim$1-2 \AA\ and $f_y
\sim$5-10\%. For early-type galaxies, this weak EW(H$\alpha$) is
mostly by post-AGB stars \cite[e.g.][]{kehrig12,papa13}, and therefore
no correlation is expected between its intensity and the fraction of
young stars (as explained before in Sec.\,\ref{HII_detect}). On the
other hand, high EW(H$\alpha$) could be produced by other mechanisms,
like AGNs and shocks, that are not required to be correlated in principle with the
properties of the underlying stellar population. A cut in the
EW(H$\alpha$) cannot remove those regions. Therefore, we consider that
the fraction of young stars provides, in connection with the
aforementioned spectroscopic classification criteria, a robust and
physically motivated means or the extraction of genuine \ion{H}{ii}
regions.

Figure \ref{fig:BPT}, left-panel, shows the distribution of the
ionized regions across the BPT diagram,  with contours indicating the
density of regions at each location. The outermost of those contours
encloses 95\% of the detected regions, with each consecutive one
encircling fewer regions. This contour is located below the
\cite{kewley01} demarcation curve, which indicates that the ionization
of our selected clumpy regions is already dominated by star
formation. In fact, only $\sim$2\% of all regions are located above
the \cite{kewley01} line, and $\sim$80\% are below the
\cite{kauffmann03} line (i.e., where classical disk \ion{H}{ii} regions
are located). If we had adopted this latter demarcation curve as our
selection criteria we would have missed a significant number of 
regions.

The color-code in Fig \ref{fig:BPT} indicates the average fraction of
young stars at each location (i.e., the $x$-axis in
Fig. \ref{fig:EW}), ranging from nearly 100\% for the regions at the
top-left area of the diagram, to nearly 0\% for regions at the
top-right location. There is a clear gradient/correlation between the
fraction of young stars and the [\ion{N}{ii}]/H$\alpha$ ratio,
reflecting the known downsizing-like variation of the specific SFR along the
SF branch of the BPT diagram \citep{asar07}.



\begin{figure*}
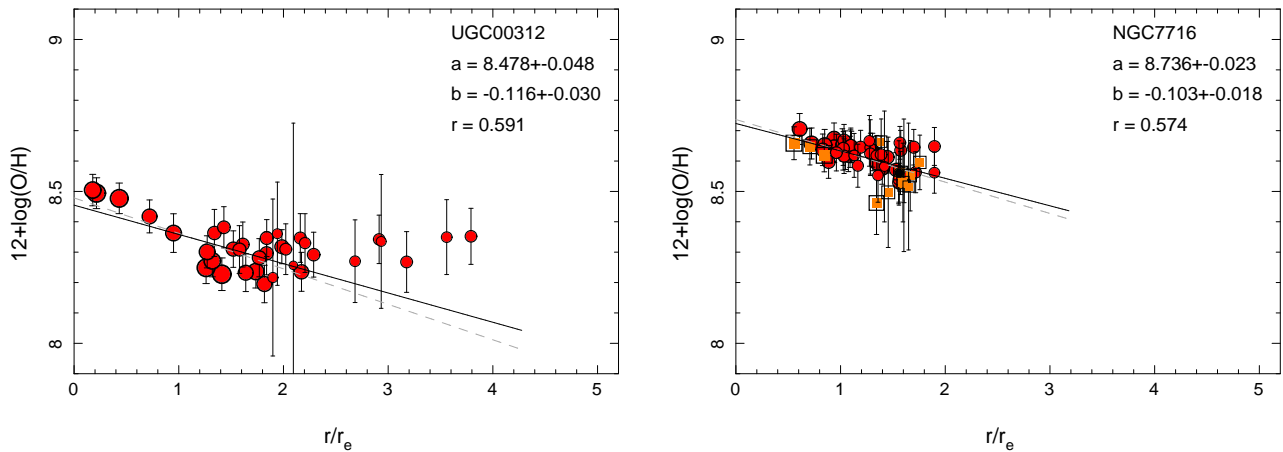

\centering
\includegraphics[width=6.1cm,angle=270,clip,trim=0 0 0 0]{figs/98UGC00312.ps}
\includegraphics[width=6.1cm,angle=270,clip,trim=0 0 0 0]{figs/98NGC7716.ps}
\caption{\label{fig:grad} Radial distribution of the oxygen abundance
  derived for the individual \ion{H}{ii} regions with abundance errors
  below 0.15 dex, as a function of the deprojected galactocentric
  distance (i.e., corrected for inclination), normalized to the disk
  effective radius, for the galaxies presented in
  Fig. \ref{fig:HII}. The size of the symbols is proportional to the
  H$\alpha$ intensity. The red circles represent those \ion{H}{ii}
  regions below the \cite{kauffmann03} line, and the orange squares
  represent those ones above this curve and below the \cite{kewley01}
  demarcation lines, i.e., the regions at the {\it so called}
  intermediate zone in the BPT diagram. The solid and dashed lines
  show the best linear regression and error-weighted linear fit
  derived for those values between 0.3 and 2.1 $r_e$. The results from
  the second fit are shown in the figure, including the zero-point
  (a), slope (b), and correlation coefficient (r). }
\end{figure*}


Based on these results, we classified as \ion{H}{ii} regions those
clumpy ionized regions for which young stars ($<$500 yr)
contribute at least a 20\% to the flux in the $V$-band. 
This particular fraction is the lowest for which the correlation coefficient
between $f_y$ and the EW(H$\alpha$) is still higher than $r_{corr}>$0.95,
and for which the fraction of excluded regions is not higher than the one 
that would be excluded by adopting the more common \cite{kauffmann03} curve.
Fig. \ref{fig:BPT}, central panel, shows the same
distribution as the one shown in the left panel, but restricted to
the 1787 regions for which the fraction of young stars is lower than 
20\%. The fraction of regions above the \cite{kewley01} curve is
significantly larger ($\sim$7\%), with more than a 40\% above the
\cite{kauffmann03} one. Although there are still 1043 regions below this
latter curve, it comprises just  $\sim$15\% of the original
sample. This can be considered our incompleteness fraction.
Although we cannot exclude that some fraction of these regions
are ionized by star-formation, we cannot guarantee it.

Fig. \ref{fig:BPT}, right panel shows the same distribution, but for
the 5229 regions with a fraction of young stars larger than 20\%,
i.e., our final sample of \ion{H}{ii} regions. Of them, only 23 are
above the \cite{kewley01} curve ($\sim$99.5\% are below it). On the
other hand, there are 713 regions in the so-called intermediate
region, with a significant fraction of young stars ($\sim$40\% on
average). These regions would have been excluded if we had adopted the
\cite{kauffmann03} curve as our selection criteria, and we would have lost a
certain number of \ion{H}{ii} regions at any galactocentric distance.  We
consider that the adopted combined selection criteria are more
physically driven and conservative, since they select only those
regions that are associated with an underlying stellar population 
indicative of the presence of young stars.

\section{Results}\label{results}

\subsection{Oxygen abundance gradients}\label{OH}


In order to derive the oxygen abundance for each of the selected
$\sim$5000 \ion{H}{ii} regions, we adopted the empirical calibrator
based on the O3N2 ratio \citep{allo79,pettini04,stas06}.

\begin{eqnarray}
 {\rm O3N2} &=& {\rm log_{\rm 10} } \left[\frac{I([\ion{O}{iii}]~\lambda5007)/I({\rm H}\beta)}{I([\ion{N}{ii}]~\lambda6584)/I({\rm H}\alpha)} \right] 
\end{eqnarray}

This ratio is basically not affected by the effects of dust
attenuation, uses emission lines covered by our wavelength range for
all the galaxies in the sample, and it has a monotonic single-valued
behavior in its range of applicability. We adopted the functional form
and calibration by \citet{pettini04}, although  its  correspondence with temperature-anchored abundances
at the high-metallicity range  is still under
debate \citep{marino13}.  In that article we
demonstrate that the indicator is valid for a range of line ratios
between $-$1.1$<$O3N2$<$1.7, which corresponds to oxygen abundances
above 12+log(O/H)$>$8 dex. In our sample of \ion{H}{ii} regions we do
not reach the low metallicity limit for which the calibration is still
useful, most probably because we do not include low-mass/dwarf
galaxies in the considered sample of galaxies. In this regime the
derived abundances have an accuracy of $\pm$0.08 dex, an uncertainty
that has been included in the error budget. The typical error derived
from the pure propagation of the errors in the measured emission lines
is about 0.05 dex, although in a few cases is can be larger.

It is beyond the scope of the current study to make a detailed
comparison of the oxygen abundances derived using the different
proposed methods, such as was presented by
\cite{kewley08} or \cite{angel12}.  However, we want to state clearly
that all our qualitative results and most of the quantitive ones are
mostly independent of the  adopted oxygen abundance
calibrator, i.e.\ despite the absolute scale among the different
indicators and the differences introduced by them in the galaxy
slopes, the abundance gradients show statistically the same
relationships with respect to global galaxy properties, as explained
below.

\begin{figure*}
\centering
\includegraphics[width=6.1cm,angle=270,clip,trim=0 0 0 0]{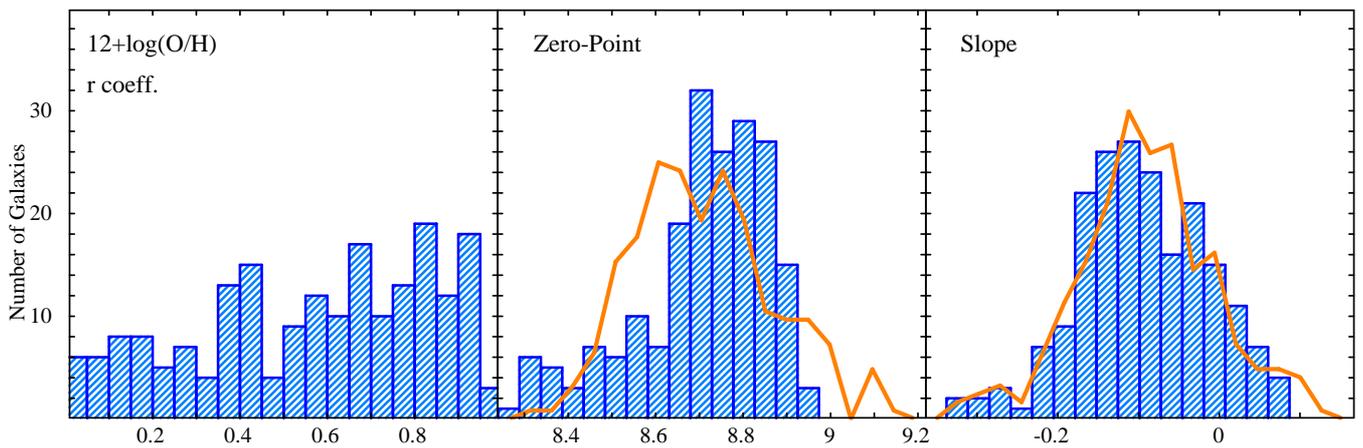}
\caption{\label{fig:stats} {\it Left panel:} Distribution of
  correlation coefficients of the oxygen abundance along the radial
  distance, for the 193 galaxies described in the text. {\it Central
    panel:} Distribution of zero-points for the corresponding linear
  regression galaxy by galaxy. {\it Right panel:} Distribution of slopes of the same
  regressions. The orange solid line represents, for each of the last
  two histograms, the expected histogram in case of a Gaussian
  distribution of the data, assuming the mean and standard-deviation
  of the distribution of each analyzed parameter, and sampled with the same bins.}
\end{figure*}

We derive for each galaxy the galactocentric radial distribution of
the oxygen abundance, based on the abundances measured for each
individual \ion{H}{ii} region. In Appendix \ref{Re} we describe
the surface-brightness and morphological analysis performed for each
galaxy to derive the mean position angle, ellipticity, and effective
radius of the disk. Using this information we deprojected the position
of each \ion{H}{ii} region for each galaxy, assuming an intrinsic ellipticity for
galaxies of $q=$0.13 \citep{giov95,giov97}, and an inclination given by:

\begin{equation}
{\rm cos^2 i = \frac{1-q^2-\epsilon^2}{1-q^2}}
\end{equation}

\noindent 

\noindent where $i$ is the inclination of the galaxy, and $\epsilon$
is the median ellipticity provided by the morphological analysis, defined 
for each galactocentric distance as:

\begin{equation}
{\rm \epsilon^2 = 1-\left(\frac{a}{b}\right)^2}
\end{equation}

\noindent where $a$ and $b$ are the semi-major and semi-minor
axes. For galaxies with an inclination below 35$\degree$ we prefer not
to correct for the inclination effects due to the uncertainties in the
derived correction, and the very small effect on the spatial
distribution of \ion{H}{ii} regions. We derive the galactocentric
distance for each region, which is later normalized to the disk
effective radius ($r_e$). This disk effective radius was derived from
the scale-length of the disk of each galaxy, extracted from the
analysis of the surface brightness profile in the $g$-band as detailed in
the Appendix \ref{Re}. For disk dominated galaxies this effective
radius is similar to the classical effective radius, that can be
derived by a pure growth-curve for the full light distribution of the
galaxy. However, for galaxies with a clear bulge it represents the
characteristic scale of only the disk part. The center of the galaxy
was taken from the WCS of the cube headers, and it was derived by a
barycenter estimation described in \cite{husemann13}.

Finally, for each galaxy we derive the oxygen abundance gradient. 
Figure~\ref{fig:grad} shows two examples of these abundance gradients for the same galaxies shown
in Figure \ref{fig:HII} (i.e.\ UGC\,00312, left-panel, with high inclination  and NGC\,7716, right-panel, with low
 inclination). As we indicated before, the
CALIFA FoV covers on average $\sim$2.5 $r_e$ of the
observed galaxies. However, due to the inclination for spiral
galaxies this FoV has a wide range between $\sim$2 $r_e$
 for the face-on galaxies and up to $\sim$5 $r_e$
 for the edge-on ones (although the particular range depends
also on the intrinsic characteristics of the galaxies).


Following this analysis we perform a linear regression, without
considering the errors of the individual abundances, and an
error-weighted linear fit to the radial distribution of abundances
galaxy-by-galaxy, restricted to the same spatial range. From the
original 227 galaxies with detected ionized regions, we restricted the
analysis to those with at least four \ion{H}{ii} regions within the
considered spatial range (0.3$<r/r_e<$2.1). Although \cite{zaritsky94}
found empirically that at least five \ion{H}{ii}
regions are required to define the slope, we found that this depends also on
the individual errors, the range of abundances and galactocentric
distances sampled, and the actual signal-to-noise. Based on a
Monte-Carlo simulation we found that for less than four
\ion{H}{ii} regions in a galaxy the derived slope is not reliable. This final
sample comprises 193 galaxies, and a total of 4610
\ion{H}{ii} regions. 94 galaxies show at least one region
beyond 2.2 disk effective radius, with a total of 484 regions (i.e,
$\sim$5 regions per galaxy in this outer region, on average).

The result of this analysis is illustrated is Fig.\,\ref{fig:stats},
where we show the distribution of the correlation coefficients,
zero-points, and slopes for each individual galaxy. For most of the
galaxies there is a clear correlation between the oxygen abundance and
the radial distance. The correlation coefficient (shown on the
left-panel of Fig.\,\ref{fig:stats}) is larger than $r_{corr}>$0.4 for
$\sim$72\% of the galaxies. This $r_{corr}$ corresponds to a
probability of good fit of $\sim$98.5\% for the typical number of
\ion{H}{ii} regions in our galaxies.  Most of the galaxies for which
the correlation coefficient is lower than this value are galaxies with
low number of detected \ion{H}{ii} regions. The distribution of
zero-points (mid-panel) has a mean value at 12+log(O/H)$\sim$8.73 dex
with a standard deviation of $\sigma\sim$0.16 dex, with a range of
values reflecting the mass-range covered by the sample, due to the
well-known \mz\ relation
\citep[e.g.][]{tremonti04,sanchez13}. Finally, the distribution of
slopes (right-panel of Fig.\,\ref{fig:stats}) has a clear peak and it
is remarkably symmetric. The probability of being compatible with a
Gaussian distribution is 98\%, based on a Lilliefors-test
\citep{tLIL67a} (compared with  77\% derived for the distribution
of zero-points). Therefore, the slopes of the abundance gradients have
a well-defined characteristic value of $\alpha_{O/H} = -$0.10
dex/$r_e$ with a standard deviation of $\sigma =$0.09 dex/$r_e$,
totally compatible with the value reported in \cite{sanchez12b}, for a
more reduced sample. This slope corresponds to an $\alpha_{O/H} =
-$0.06 dex/$r_d$, when normalized to the disk scale-length ($r_d$),
instead of the disk effective radius ($r_e$).  If instead of this
normalization scale, we adopt a more classical one, like
$r_{25}$ (the radius at which the surface-brightness reaches 25
mag/arcsec$^2$ in the B-band) we obtain a similar result, although for
a sharper slope of $\alpha_{O/H} = -$0.16 dex/$r_{25}$, and a dispersion
of $\sigma=$0.12 dex/$r_{25}$. Finally, if the physical scale  (i.e., kpc) at the
distance of the galaxy is used instead of any of the previous
normalizations, then we find a shallower average slope of
$\alpha_{O/H} = -$0.03 dex/kpc with a standard deviation of
$\sigma=$0.03 dex/kpc. Even more important, for this final case
the distribution is not asymmetric, presenting a clear tail towards
large slopes, up to $-$0.15 dex/kpc.



\subsection{Abundance gradient by galaxy types}\label{morph}



\begin{figure*}
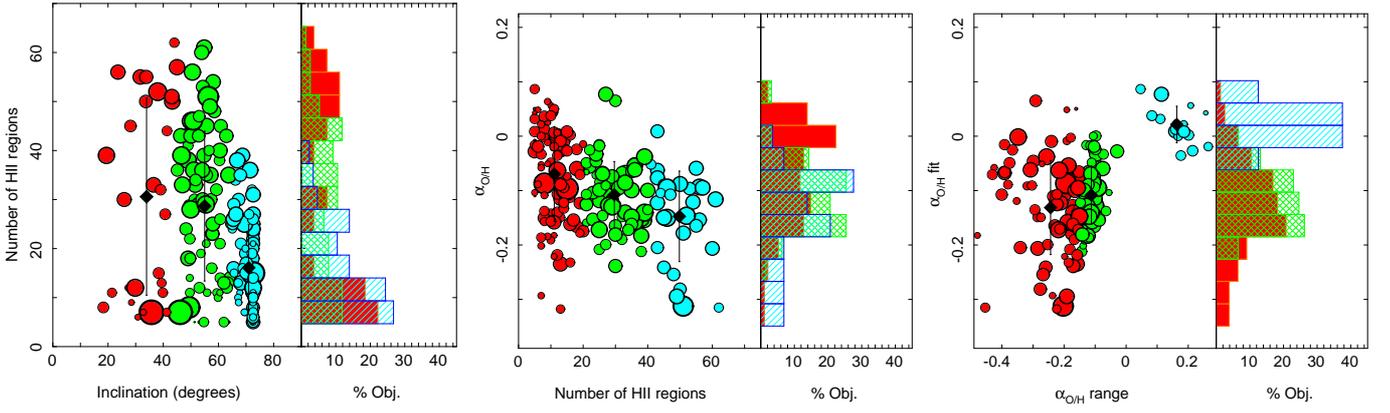

\centering
\includegraphics[width=5.5cm,angle=270,clip,trim=0 0 0 0]{figs/N76Mass_OH_3.ps}
\includegraphics[width=5.5cm,angle=270,clip,trim=0 0 0 0]{figs/56Mass_OH_3.ps}
\includegraphics[width=5.5cm,angle=270,clip,trim=0 0 0 0]{figs/5066Mass_OH_3.ps}
\caption{\label{fig:bias}{\it Left panel:} Number of detected \ion{H}{ii}
  regions as a function of the inclination of the galaxy. The colors of the
  symbols and the corresponding histogram indicate the inclination of the
  galaxies: (i) Less than 45$\degree$ (red), (ii) between 45$\degree$ and
  65$\degree$ (green) and (iii) larger than 65$\degree$ (blue). {\it Central
    panel:} Slope of the gradients of the oxygen abundance derived for each
  galaxy along the number of detected \ion{H}{ii} regions. The colors of the
  symbols and the corresponding histogram indicate the number of detected \HII
  regions in each galaxy: (i), Less than 20 \ion{H}{ii} regions (red), (ii)
  between 20 and 40 \ion{H}{ii} regions (green) and (iii) more than 40 \HII
  regions (blue). {\it Right panel:} Slope of the gradients of the oxygen
  abundance derived for each galaxy based on the linear regression of the
  radial distribution along the slope derived by comparing the range of
  abundances with the corresponding range of radial distances. The colors of
  the symbols and the corresponding histogram indicate this latter parameter,
  showing (i) slopes lower than than $-$0.25 dex/$r_e$ (red), (ii) between
  $-$0.25 and 0.05 dex/$r_e$ (green) and (iii) larger than 0.05 dex/$r_e$
  (blue). The size of the symbols are inversely proportional to the derived
  error in the slope of the abundance gradient, in both panels. The
  black-solid diamonds represent the mean values for the different selected
  subsamples, with the error bars indicating the standard deviation around this
  mean value. }
\end{figure*}

In this section we analyze the possible dependence of the slope of
the gradients on the properties of the galaxies. But
before addressing this issue we point out some possible
limitations and biases affecting the analysis performed. 
Figure \ref{fig:bias}, left panel, shows the distribution of the number of
detected \ion{H}{ii} regions as a function of the inclination of the
galaxy. It is clear that although the number of \ion{H}{ii} regions is not the
only parameter that affects this error, for galaxies with less than $\sim$10
regions the error is considerably larger. On the other hand the number
of detected regions decreases with increasing inclination. For highly
inclined galaxies ($i>70\degree$), there are very few galaxies with
more than 15 \ion{H}{ii} regions. This is a clear selection effect,
since highly inclined galaxies have less accessible portion of the
disk, and therefore the number of detected regions is reduced. We have taken
into account this bias in the following analysis.

Fig \ref{fig:bias}, central panel, shows the distribution of the slopes as a
function of the number of detected \ion{H}{ii} regions, including galaxies of
any inclination. For galaxies with few detected regions there is a strong
secondary peak in the distribution at $\alpha_{O/H}\sim$0 (i.e., a constant
value). This secondary peak is more evident in the right panel, where we
compare the slopes derived from the linear regression (i.e., those shown in
the central panel and in Fig. \ref{fig:stats}), with a rough estimation of the
slope derived by dividing the range of abundances within the considered
galactocentric distances ($0.3-2.1$ $r_e$), by the differences of radial
distances,

\begin{equation}
{\rm \alpha_{range} = \frac{max[12+log(O/H)]-min[12+log(O/H)]}{r_{max\ O/H} - r_{min\ O/H}}}.
\end{equation}

\noindent This parameter is more sensitive to the actual range of
abundances measured for the \ion{H}{ii} regions in each galaxy. Most
of the galaxies are concentrated in a cloud around
$(\alpha_{O/H},\alpha_{range})=(-0.1,-0.15)$ (with a wide dispersion
in the second parameter). However, there is a second group of galaxies
with nearly flat or even inverse gradients (indicated with a blue
color), which are mostly galaxies with a low number of \ion{H}{ii}
regions and/or highly inclined galaxies. It is clear that for those
galaxies our derived slope is less reliable. Thus, better determinations of the slope will result for  (i) larger number
of \ion{H}{ii} regions, (ii) larger range of abundances, and (iii) larger covered
range of galactocentric distances.

\begin{figure*}
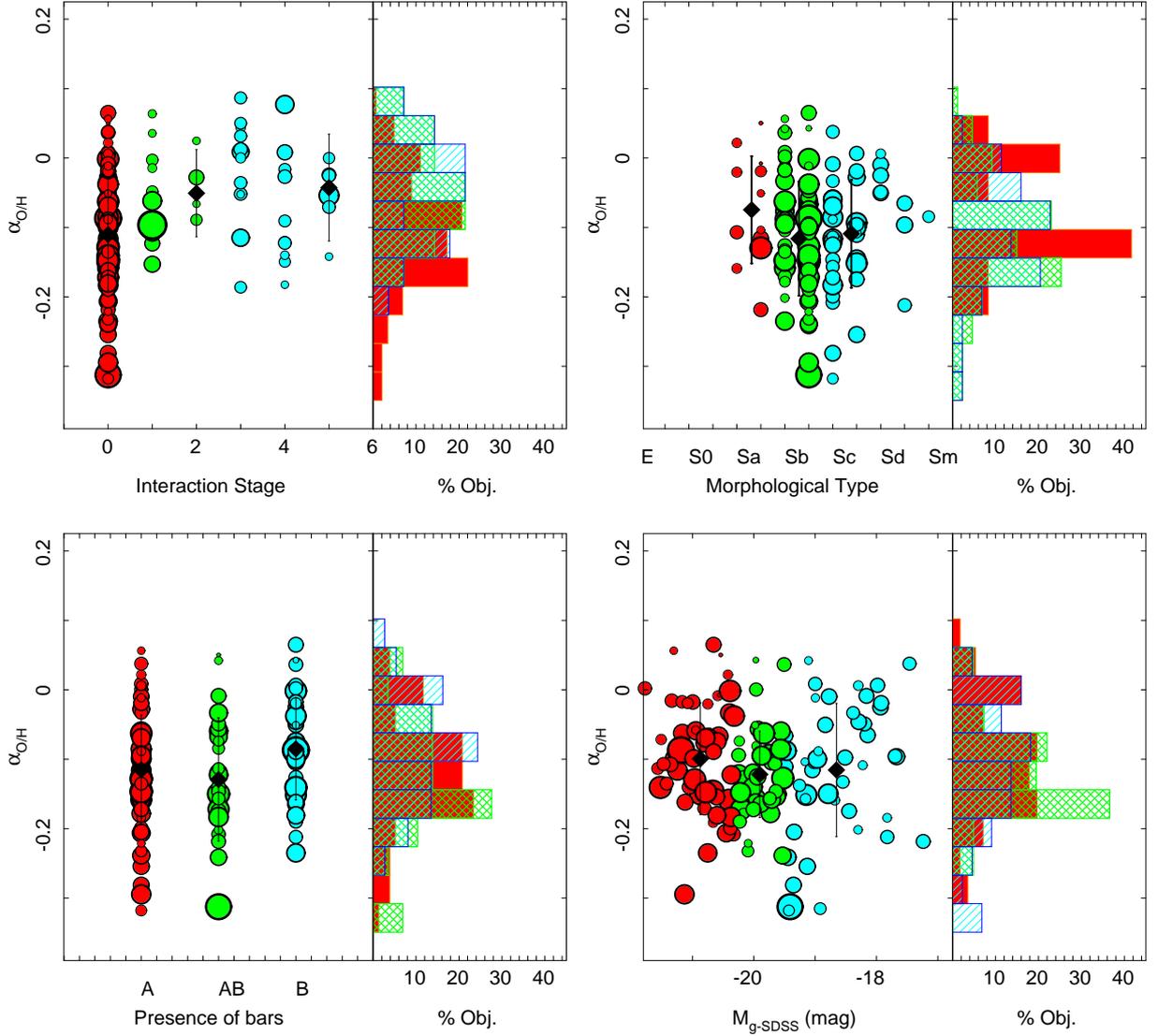

\centering
\includegraphics[width=7.5cm,angle=270,clip,trim=0 0 0 0]{figs/M6Mass_OH_3.ps}
\includegraphics[width=7.5cm,angle=270,clip,trim=0 0 0 0]{figs/6Mass_OH_no_inter.ps}
\includegraphics[width=7.5cm,angle=270,clip,trim=0 0 0 0]{figs/16Mass_OH_no_inter.ps}
\includegraphics[width=7.5cm,angle=270,clip,trim=0 0 0 0]{figs/36Mass_OH_no_inter.ps}
\caption{\label{fig:morph}{\it Top-left panel:} Distribution of the slopes of
  the abundance gradients as a function of the interaction stage of the
  galaxies. The colors of the symbols and the corresponding histograms
  indicate three types of galaxies based on the interaction: (i) no signatures
  of interaction (red), (ii) galaxies with close companions and/or in an early
  interaction stage (green), and (iii) galaxies under clear coalescence or
  evolved mergers (blue). {\it Top-right panel:} Similar distribution of
  slopes as a function of the morphological classification of the
  galaxies. The colors of the symbols and the corresponding histograms
  indicate three types of galaxies based on their morphology: (i) Early
  spirals, SO-Sa (red), (ii) intermediate spirals, Sab-Sb (green) and (iii)
  Late spirals, Sc-Sm (blue). {\it Bottom-left panel:} Similar distribution
  of slopes as a function of the presence or absence of bars. The colors of
  the symbols and the corresponding histograms indicate three types of
  galaxies: (i) clearly non-barred (red), (ii) not clear if there is a bar or
  not (green) and (iii) clearly barred galaxies (blue). {\it Bottom-right
    panel:} Similar distribution of slopes as a function of the absolute
  magnitude of the galaxies. The colors of the symbols and the corresponding
  histograms indicate three types of galaxies based on the luminosity: (i)
  luminous galaxies, $M_{g-SDSS}<-20.25$ mag (red), (ii) intermediate galaxies
  $-19.5<M_{g-SDSS}<-20.25$ mag (green) and (iii) faint galaxies
  $M_{g-SDSS}>-19.5$ mag (blue). The size of the symbols are inversely
  proportional to the derived error in the slope for all the panels . The
  black-solid diamonds represent the mean values for the different selected
  subsamples, with the error bars indicating the standard deviation around this
  mean value. }
\end{figure*}

\subsubsection{Effects of interactions in the abundance gradients}\label{slope_inter}


We classified our sample of galaxies based on their interaction stages
to study the possible effect in the abundance gradient, with a much
stronger statistical basis than any previous study. Following the
classification scheme by \cite{veil95}, galaxies were classified in
six different groups, from (i) galaxies without any evidence of
interaction (class 0), like NGC\,5947; (ii) galaxies with close
companions at similar redshift (classes 1-2), like VV\,448; and (ii)
galaxies under clear interaction and/or advanced mergers (classes
3-5), including galaxies like the Mice (class 3) and ARP\,220 (class
4). The details of these classification will be given elsewhere
(Barrera-Ballesteros et al., in prep.).

Figure \ref{fig:morph}, top-left panel, shows the distribution of
slopes of the abundance gradient for the different classes based on
the interaction stage. Most of the galaxies in this study do not present any
evidence of an on-going interaction ($\sim$77\%). But they do present a well
centred distribution of slopes, with an average value of
$\alpha_{O/H}=-$0.11 dex/$r_e$ with a standard deviation of $\sigma=$0.08
dex/$r_e$, fully compatible with the distribution for the complete sample
(based on a Kolmogorov-Smirnov test, hereafter KS-test). On the other hand,
the two subsamples of galaxies with evidence for early or advanced
interactions present similar distributions of slopes among themselves, with
shallower gradients ($\alpha_{O/H}=-$0.05 dex/$r_e$ and $\sigma=$0.07
dex/$r_e$), significantly different from the subsample of
non-interacting galaxies: $p_{KS}=$96.19\% for classes 1-2 and
$p_{KS}=$99.96\% for classes 3-5. It is clear that the disk effective
radius is most probably ill defined for advanced mergers, however,
this is irrelevant when the slope is close to zero.
Figure \ref{fig:bias}, right panel, shows that most of the galaxies with
a flat slope are galaxies with a narrow range of abundances across the
field-of-view, and those ones are mostly interacting/merging galaxies.
Therefore, we conclude that galaxy interactions flatten the abundance
gradient.

Moreover, we restricted our analysis for the 106 galaxies with more
than 10 \ion{H}{ii} regions, and with inclinations lower than
70$\degree$, taking into account the possible biases described in the
previous section. We found no qualitative difference in the result.
For the intermediate stage the actual number of galaxies is too low
(7) to provide with a significant difference (although the mean value
of the slope remains the same). Finally, for the advance mergers the
difference in slope remains significant ($p_{KS}=$99.81\%).

\subsubsection{Slopes by morphology}\label{slope_morph}


\begin{figure*}
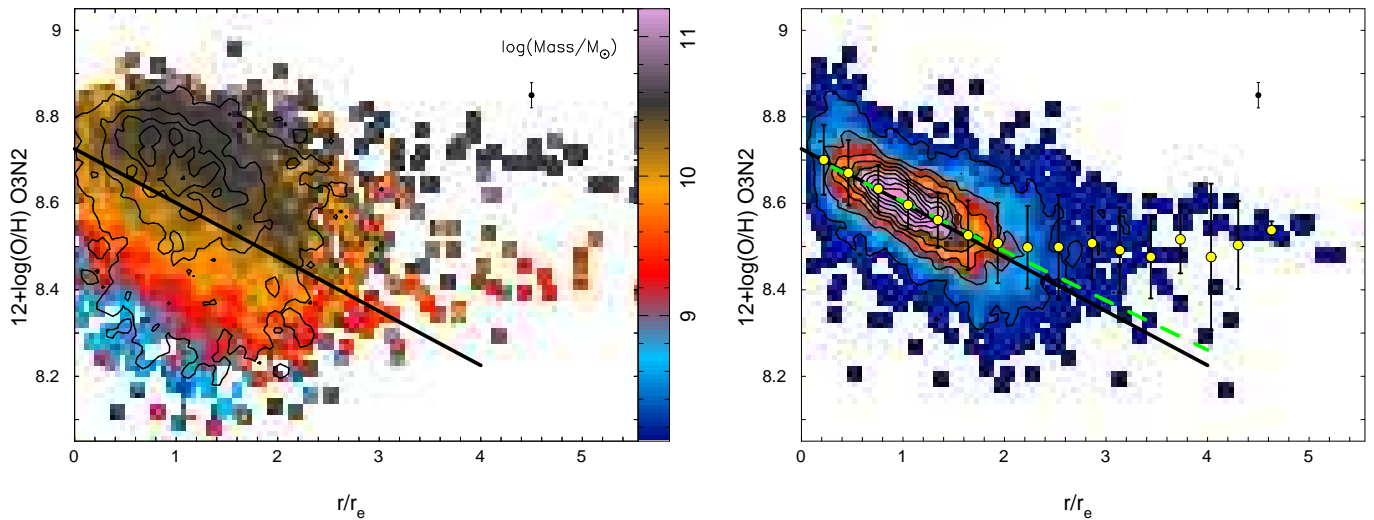

\centering
\includegraphics[width=7.cm,angle=270,clip,trim=0 0 0 0]{figs/Re_OH_dep_NO_SCALE_Mass.ps}
\includegraphics[width=7.cm,angle=270,clip,trim=0 0 0 0]{figs/Re_OH_dep.ps}
\caption{\label{fig:grad_OH} {\it Left panel:} Radial distribution for the oxygen
  abundance derived using the O3N2 indicator for all the galaxies in our sample. The
  contours show the density distribution of \ion{H}{ii} regions in
  this parameter space. The outermost contour encircles 95\% of the total
  number of \HII regions, with $\sim$20\% less in each consecutive
  contour. The color image shows the average stellar mass of each galaxy
  corresponding to each abundance and radial distance. {\it Right panel:}
  Radial distribution for the oxygen abundance derived using the O3N2
  indicator, after scaling to the average value at the disk effective radius
  for each galaxy. The image and contours show the density distribution of
  \ion{H}{ii} regions in this parameter space. The outermost contour encircles
  95\% of the total number of \HII regions, with $\sim$20\% less in each
  consecutive contour. The solid-yellow points represent the average oxygen
  abundances, with their corresponding standard deviations indicated
  as error bars for consecutive bins of 0.3 galactocentric distances
  per disk effective radius. The average error of the derived oxygen
  abundance (without considering systematic errors) is shown by a
  single error bar located a the top-right side of the panel. The dashed-green
  line shows the result of the best linear regression to the data. The
  solid-black lines in both panels represent the linear relation corresponding
  to the mean values of the zero-points and slopes of the individual
  regressions derived for distribution of each individual galaxy.}
\end{figure*}

In Fig. \ref{fig:morph}, top-right panel, we show the distribution of
slopes as a function of the galaxy morphological classification. This
classification was performed by eye, based on the independent analysis
by five members of the CALIFA collaboration, and it will be described
elsewhere in detail (Walcher et al. in prep.). Different tests indicate
that our morphological classification is fully compatible with
pre-existing ones, and their results agree with the expectations based
on other photometric/morphological parameters, like the concentration
index (Fig \ref{fig:CM}) and/or the S\`ersic index \citep{sersic68}.
We exclude from this analysis those galaxies with evidence of an on-going
interaction (i.e., classes 1-5, in the previous section), since they
present a much flatter gradient. This reduces our sample to 146 galaxies.

The earlier spirals (S0/Sa) present a slightly flatter slope,
$\alpha_{O/H,S0-Sa}=-$0.08 dex/$r_e$ and $\sigma=$0.08 dex/$r_e$ ($n_{gal,S0-Sa}=$13), in
comparison with the two groups of later type ones:
$\alpha_{O/H,Sab-Sb}=-$0.12 dex/$r_e$ and $\sigma=$0.08 ($n_{gal,Sab-Sb}=$88) dex/$r_e$ and
$\alpha_{O/H,Sbc-Sm}=-$0.11 dex/$r_e$ and $\sigma=$0.08 dex/$r_e$
($n_{gal,Sbc-Sm}=$45). However, the corresponding $t$- and KS-tests
indicate that the differences are not significant: $p_{t}=$98.86\% and
$p_{KS}=$81.31\%, respectively, for the distributions with the larger
differences. Therefore, statistically speaking, the slopes of spirals galaxies
segregated by morphology are all equivalent.

A similar result is found if instead of normalizing the radial
distances to the disk effective radius we adopt the physical size,
without any scale-length normalization. In this case we derive a much
wider distribution of slopes, not compatible with a Gaussian
distribution (as indicated before). The only difference is that the
values for the early-type galaxies are somehow shallower and with a
narrower range than that of the  later types, although the
differences are  not statistically significant.

\subsubsection{Effects of bars in the abundance gradients}\label{slope_bar}


Fig. \ref{fig:morph}, bottom-left panel, shows the distribution of
slopes for the different types of galaxies, depending on the clear presence,
or not, of bars. The inspection of our sample for bars was performed by eye,
by five members of the CALIFA
collaboration, and it will be described elsewhere in detail (Walcher
et al. in prep.). Three different groups were defined, following the
classical scheme: (A) galaxies with no bar, (AB) galaxies that may
have a bar, but it is not clearly visible and (B) clearly barred
galaxies. The visual classification was cross-checked with an
automatic search for bars, based on the change of ellipticity and PA,
that has yield similar results. In a recent kinematical analysis
of the H$\alpha$ velocity maps, using {\tt DiskFit}\footnote{http://www.physics.rutgers.edu/~spekkens/diskfit/},
it was found that the frequency of radial motions was significantly higher in those
galaxies with clear bars (Holmes et al., in prep.). As for the previous
section, we only considered the 146 galaxies with no evidence of an on-going
interaction.

Again, negligible differences in statistical terms are found between the slope
of the abundance gradient for barred galaxies, i.e. $\alpha_{O/H,B}=-$0.09
dex/$r_e$ and $\sigma=$0.07 dex/$r_e$ ($n_{gal,S0-Sa}=$38), in comparison with
the other two groups: $\alpha_{O/H,A}=-$0.12 dex/$r_e$ and $\sigma=$0.08
($n_{gal,A}=$78) dex/$r_e$ and $\alpha_{O/H,AB}=-$0.13 dex/$r_e$ and
$\sigma=$0.09 dex/$r_e$ ($n_{gal,AB}=$30). The corresponding $t$- and KS-tests
indicate that the differences are not significant: $p_{t}=$93.43\% and
$p_{KS}=$92.44\%, respectively, for the distributions with the largest
differences. Therefore, if there is a change in the general slope of the
gradient induced by the presence of a bar, the effect is weak and not
statistically significant. The same result is found if the radial distances
are normalized to the physical size, without any scale-length
normalization. As in the previous case, we derive a wider distribution of
slopes, but with no significant differences due to the presence or absence of
bars.

\subsubsection{Slopes by luminosity, stellar mass and concentration index}\label{slope_lum}

In the previous sections, we analysed the possible changes of the
slope of the abundance gradient on the basis of three different
morphological classifications: merging/interaction stage, Hubble type
and incidence of bars. All those classifications were performed by
eye, deriving a discrete segregation of the galaxies in sub-groups. 
In this section we analyse the possible variation of the slope as a function of
less-subjective parameters, that are correlated with the
morphology of the galaxies: the luminosity, the stellar mass and the
concentration index $C$ (see definition in Sec. \ref{sample}).


Fig. \ref{fig:morph}, bottom-right panel, illustrates this analysis,
showing the distribution of slopes versus the $g$-band absolute
magnitude of the galaxies. As in the previous sections we excluded
the galaxies with clear interactions. We split the sample in luminous
($L$, M$_g<=-$20.25 mag), intermediate ($I$, $-$20.25$<$M$_g<=-$19.5 mag),
and faint ($F$, M$_g>-$19.5 mag) galaxies. No significant difference is
found between both the average slopes: $\alpha_{O/H,L}=-$0.10 dex/$r_e$ and $\sigma=$0.08
dex/$r_e$ ($n_{gal,L}=$57), $\alpha_{O/H,I}=-$0.12 dex/$r_e$ and $\sigma=$0.06 dex/$r_e$
($n_{gal,I}=$42), $\alpha_{O/H,F}=-$0.12 dex/$r_e$ and $\sigma=$0.10 dex/$r_e$
($n_{gal,F}=$45). The corresponding $t$- and KS-tests indicate that
the probability that they are different are $p_{t}=$88.11\% and
$p_{KS}=$87.36\%, respectively, for the sub-samples with the largest
differences. There is not even a weak trend between both
parameters, given the derived correlation coefficient $r=$0.009, and a
slope provided by a linear regression of $a=-$0.0007. Thus the abundance
gradient seems to be independent of the luminosity of the galaxies.

Similar results are found for the stellar masses
\citep[derived as described in][]{sanchez13}, and the $C$-index. No clear
correlation is found between the slopes of the abundance gradients and both
parameters, as indicated by the derived correlation coefficients:
$r_{mass}=$0.08 and $r_{C-index}=$0.24. The only difference is found
for galaxies more massive than 4.5$\times$10$^{10}$ M$_{\odot}$, with
concentration indices larger than $C>$2.4 (34 galaxies). These
galaxies present an average slope of $\alpha_{O/H}=-$0.07 dex/$r_e$ and $\sigma=$0.06
dex/$r_e$, and $t-$ and KS-tests indicate that they are significantly
different from the rest of the sample: $p_{t}=$99.60\% and
$p_{KS}=$99.08\%. However, even this difference has to be taken with
care, since a visual inspection of the abundance gradients for this
subsample indicates that a substantial fraction of them are galaxies
with few detected \ion{H}{ii} regions.

\section{Discussion}\label{discuss}

The metal content of a galaxy is a fundamental parameter to understand
the evolution of the stellar populations galaxy-by-galaxy and at
different locations within the same galaxy. Oxygen is the most
abundant heavy element in the Universe, making it the best proxy of total
metallicity. It is easily observable for a wide range of metallicities
thanks to its emissivity of collisionally excited lines, which are
prominent in the optical regime.  The existence of an universal radial
decrease in the oxygen abundance has been already suggested in many
previous studies
\citep[e.g.][]{diaz89,VilaCostas:1992p322,bresolin09,yoac10,rosales11,mari11,sanchez12b,bresolin12}.
This observational property is compatible with our current
understanding of the formation and evolution of spiral galaxies
\citep[e.g.][and references therein]{tsuj10}. Gas accretion brings
gas into the inner region, where it first reaches the required density
to ignite star formation. Thus the inner regions are populated by
older stars, and they have suffered a faster gas reprocessing, and
galaxies experience an inside-out mass growth
\citep[e.g.][]{matt89,bois99}.  Several previous studies have analyzed
the radial abundance gradients for individual galaxies or for limited
samples of galaxies
\citep[e.g.,][]{VilaCostas:1992p322,belley92,zaritsky94,roy97,Zee1998,mari11,rosales11,rich12,bresolin12}.
These studies have found: (i) a monotonic decrease of the abundance
from the central regions, up to $r_{25}$ and/or $\sim$2.5 -- 4 $r_0$ (the
scale-length of the disk), which corresponds basically to
$\sim$1.5 -- 2.5 $r_e$; (ii) a flattening in the outer regions, for those
galaxies that cover regions beyond $r_{25}$; (iii) in some cases a
shallow drop of the abundance in the central regions is found
\citep[e.g.][$<$0.3-0.5 $r_e$]{rosales11}.  In \cite{sanchez12b} we
presented the first study of a large number ($\sim$2000) of \ion{H}{ii}
regions, extracted from 38 face-on spiral galaxies. In general, we
confirmed the common pattern described above (although the sample of
regions beyond $\sim$2$r_e$ was quite reduced), we found that there is
not only a common pattern but a common slope of $\alpha_{O/H}= -$0.12
dex/$r_e$ for all the abundance gradients between 0.3-2.1$r_e$, when
normalized to the disk effective radius of the galaxies.

The results presented in the previous section point to the same
conclusion as  \citet{sanchez12b}, i.e., that independently of the
large variety of analyzed galaxies, {\em disk galaxies in the local
  Universe present a common/characteristic gradient in the oxygen
  abundance up to $\sim$2 disk effective radii}. Moreover, the
distribution around this mean value is compatible with a Gaussian
function, and therefore could be the result of random fluctuations.
This result contradicts several previous studies which claim that the
slope in the gas-phase abundance gradient is related to other
properties of the galaxies, such as (i) the morphology, with early-type
spirals showing a shallower slope and late-type ones a sharper one
\citep[e.g.][]{McCa85,VilaCostas:1992p322}, (ii) the mass, with more
massive spirals showing a shallower slope and less massive ones a
sharper one \citep[e.g.][]{zaritsky94,Martin:1994p1602,garnet98}; and
in particular, (iii) the presence of a bar, with barred galaxies
presenting a shallower slope than non-barred ones
\citep[e.g.][]{zaritsky94,roy96}, and (iv) the interaction stage of
the galaxies, with evolved mergers presenting shallower slopes
\citep[e.g.][]{rich12}, which seems to be the case also for irregular
galaxies and low-mass galaxies
\citep[e.g.][]{edmunds93,walsh97,kobu97,molla99,kehrig08}.

In the first case, the dependence of oxygen abundance on the morphology of the
galaxies is a long standing debate \citep[e.g.][]{McCa85,VilaCostas:1992p322}.
Early results indicated that early-type spirals (S0/Sa) present flatter gradients
than late-type ones (Sc/Sm), although these results were based on a handful of
observed galaxies and it was never tested in an statistical sense, until the
study presented here. Nevertheless, as described in Sec.~\ref{slope_morph},
statistically speaking, the slopes of spirals galaxies segregated by
morphology are all equivalent.
In the case of the bars, it is well-known that at least a 30\% of disk
galaxies have a pronounced central bar feature in the disk plane and many more
have weaker features of a similar kind \citep[e.g.][]{sell93}. Kinematic data
indicate that the bar constitutes a major non-axisymmetric component of the mass
distribution which  tumbles rapidly about the axis normal to
the disk plane. The theory predicts that bars are only stable inside
co-rotation, although whether they are stable or not remains under discussion
\citep[e.g.][]{jogee04,mendez12}. The bar and the spiral arms present two
separate pattern speeds, with the bar rotating much faster, as has been
recently observed \citep[e.g.][]{iperez12}.

Bars have been proposed as an effective mechanism for radial migration
\citep[e.g.][]{dimat13}. Hydrodynamical simulations have shown that
bars induce angular momentum transfer via gravitational torques, that
result in radial flows and mixing of both stars and gas
\citep[e.g.][]{atha92}. This radial movement can produce a mixing and
homogenization of the gas, which leads to a flattening of any
abundance gradient \citep[e.g.][]{fried94,frie98}. Resonance patterns between
the bar and the spiral pattern speed can shift the orbits of stars,
mostly towards the outer regions \citep{minc10}, a mechanism that
affects also the gas. Another process that produces a similar effect is
the coupling between the pattern speed of the spiral arms and the bar
that induces angular momentum transfer at the co-rotation radius
\citep[e.g.][]{sellwood02}. Early observational results described a
flattening in the abundance gradient of barred galaxies
\citep{zaritsky94,Martin:1994p1602}.  However, as first described in
\citet{sanchez12b}, and explained in Sec.~\ref{slope_bar} of this
work, we found negligible differences in statistical terms between the
slope of the abundance gradient for barred galaxies, i.e., no evidence
of the claimed flattening. 

A direct comparison between our derived slopes and those presented by
previous results is dangerous, due to the inhomogeneity of the
data. However, it is needed to investigate the source of the
discrepancies. \citet{zaritsky94} presented an analysis of the
abundance gradient based on a sample of 39 galaxies. Of them, 14 were
new objects (comprising a total of 159 \ion{H}{ii} regions), and the
remaining ones were extracted from the literature. Finally, only 7
objects of the total sample present a clear bar. Although in general
their abundance estimations cover a radial range up to $\sim$2
effective radii (or one isophotal radius in their nomenclature), in
many cases the actual sampled spatial range is much more reduced (eg.,
NGC~1068 or NGC~4725, as can be seen in their Fig. 8). In other cases
the slope is derived from a very low number of \ion{H}{ii} regions (in
particular some of the largest derived slopes), with values that have
recently been updated to much smaller values \citep[e.g., NGC~628, for
  which they derive a slope of $-$0.96 dex/$r_{25}$, when the actual
  value is 5 times
  lower,][]{sanchez11,rosales11,sanchez12b}. Therefore the individual
measurements presented in this article have to be taken with care,
although they were most probably the best available at the
time. Finally, the claim that barred galaxies present a flatter
abundance gradient than non-barred ones is based on the comparison of
this parameter for 7 barred with 32 non-barred galaxies (Fig.15 of
that article). The strongest difference is found in the very late-type
galaxies where there are just a few objects (3 barred and 3 non-barred
galaxies, with Hubble Type T$>$6). Some of the reported values are
hardly feasible, since slopes of $-$0.75 dex/$r_{25}$ could imply a
range of abundances that is not observed across a galaxy in more
recent estimations. On the other hand no distinction was made between
interacting and non-interacting galaxies, which may introduce another
source of uncertainty, since those galaxies present a flatter
distribution \citep[e.g.][, and this study]{rupk10}.

\cite{Martin:1994p1602} presented a comparison of the abundance
gradient based mostly on literature data. Of the 24 analyzed galaxies,
they present new data for three barred galaxies, two of them with an
apparent flatter abundance gradient than non-barred galaxies of the
same morphological type (NGC~925 and NGC~1073). However, the slope they
presented for those galaxies, when normalized by the effective radius,
cannot be considered flatter than the average ($-$0.185 dex/$r_e$ and
-0.254 dex/$R_e$, respectively, extracted from Table 7A,B from that
article). Only when they compare the gradients of the barred and
non-barred galaxies normalized to the physical scale (dex/kpc) can they
find a difference, although they advise that "the sample of each
morphological type is small".  We have already indicated along this
study the importance of defining the gradient normalized to the
effective radius, since this parameter presents a clear correlation
with other parameters of the galaxies, such as the absolute magnitude,
the mass and the morphological type.

Their analysed sample comprises a heterogeneous selection of objects
with the main criteria that they have at least 10 \ion{H}{ii} regions
with published abundance estimates in each galaxy. Of the 24
galaxies, nine have a large inclination angle ($>$55$\degree$,
including NGC~925), and no inclination correction has been applied in
the derivation of the abundance gradient. This may have an impact on
the derived slopes. Finally, the three galaxies with flatter gradients
($=0$ dex/kpc) have been classified as merging systems, and two of
them have been classified as barred galaxies. On the other hand, the
galaxy with stronger gradient is also a merging system, being
classfied as non-barred. If both highly inclined objects and merging
systems are excluded from the comparison, there is no clear evidence
of a difference between barred and unbarred galaxies in their sample.

The same result is found for the possible variation of the slope as a
function of galaxy properties which are correlated with the morphology
of the galaxies, i.e. the luminosity, the stellar mass and the
concentration index $C$, as described in Sec.~\ref{slope_lum}. When
compared with previous literature data that described correlations
with these parameters, we have to take into account how these results
were derived \cite[e.g.][]{zaritsky94,garnet98}, i.e., the analyzed
sample of galaxies and \ion{H}{ii} regions, and if they refer to the
slope normalized to the physical scale or to a certain galactic
scale-length. For example, \cite{zaritsky94}, despite  the caveats
expressed before about their sample, found a correlation between the
slope and different properties of the galaxies, such as the stellar mass,
the rotation velocity and the morphological type, but only when the
slope was expressed in terms of the physical size (dex/kpc). Since it
is well-known that the effective radius of a galaxy correlates with
those parameters, both results may be compatible, as suggested already
in \cite{sanchez12b}.

Finally, it is important to remember that in many cases well established
results are based on reduced and heterogenous samples of galaxies with
an insufficient number of \ion{H}{ii} regions explored, or with
comparison samples also extracted from literature data. \cite{berg13}
explain that most  old estimates of radial gradient of oxygen
abundances in the literature are based on out-of-date strong-line
empirical calibrations which may be uncertain. These authors measure
the auroral line [\ion{O}{iii}]$\lambda$4363 and find a gradient of
-0.017 dex/kpc and -0.027 dex/kpc for NGC~638 and NGC~2403,
respectively, which correspond to -0.10 and -0.15 dex/$r_e$ in perfect
agreement with results. These values are much smaller than the old
ones, as also occurs with the estimates from \cite{bresolin11} and
\cite{bresolin12} for M~33, M~31, NGC~4258 and M~51, which are now -0.042,
-0.023, -0.011 and -0.020 dex/kpc, smaller than the old numbers
reported by \cite{zaritsky94}. In fact the value of -0.017 dex/kpc for
NGC~628 is very similar to the one found by \cite{rosales11}, as
indicated before \citep[compatible with the value reported
  by][]{berg13}. Therefore the procedure to measure the gradient is
important, as well as the number of points or other important
observational effects, such as the angular resolution, the signal to
noise, or the annular binning that may also change the obtained radial
gradient, such as it is demonstrated in \cite{yuan11} and Mast et
al. (in prep.).

In the case of the effects of interactions on the abundance gradient of
galaxies, theory predicts that major mergers trigger the formation of bars in
the stellar and gas disks, which induce vigorous gas inflows as the
gas looses angular momentum to the stellar component \citep{barn96}.
These in-flows are thought to be responsible for fueling a massive
central starburst and feeding AGN and/or quasar activity
\citep{mihos96,barn96}. For a spiral galaxy with a preexisting
metallicity gradient, gas in-flow flattens the gradient by diluting the
higher abundance gas in the central regions with the lower abundance
gas from the outer parts \citep{rupk10,rupk10b,kewley10}. This flattening is
compounded as the spiral arms are stretched by tidal effects
\citep{torr12}. In addition, interactions induce central star formation
processes that produce violent outflows that eject metals from the richer
central regions (e.g., Wild et al., in prep.). Indeed, galaxy mergers and
interacting systems seem to present flatter gradients in the
oxygen abundance \citep[e.g.][]{kewley10,rich12}.
The results of Sec.~\ref{slope_inter} support the same scenario, i.e., that
galaxy interactions flatten the abundance gradient of spiral galaxies in a
statistical sense.

\subsection{The common abundance gradient}\label{common}

The results of the present study indicate that, when using
the O3N2 strong-line abundance indicator, the oxygen abundance gradient in
disk galaxies present, {\em statistically}, a common slope of $\sim-$0.1 dex/$r_e$,
between 0.3 and 2.1 disk effective radius, when normalized to this disk
effective radius. This common slope is independent of the other properties of
the galaxies, except for interaction/merging and maybe for the more massive
and concentrated galaxies. 

Following \cite{sanchez12b} it is possible to illustrate this result
by presenting the radial distribution of the oxygen abundance for all
the galaxies in a single figure. Since the representative abundance
(i.e., the abundance at the disk effective radius) scales with the
integrated mass \citep{sanchez13}, following the well known
\mz\ relation \citep{tremonti04}, it is required to apply a global
offset galaxy by galaxy, and normalize the gradient by the mean value
at the disk effective radius for all the sample, i.e.,
12\,+\,log(O/H)\,=\,8.6 dex. Figure \ref{fig:grad_OH} illustrates this
process. In the left panel, we show the contour-plot  of the
radial distribution before re-scaling to the average representative
abundance. The distribution comprises $\sim$4500 \ion{H}{ii} regions
corresponding to 193 different galaxies of any morphological type,
including barred and un-barred galaxies, and covering all the
CM-diagram to the completeness limit of the CALIFA survey (described
in Walcher et al. in prep.). Although the radial gradient is still
visible, there is a wide distribution that almost blurs the evidence of
a common abundance gradient. The color-coded areas represents the
integrated stellar mass of each galaxy corresponding to each
represented abundance (in log units). As expected from the
\mz\ relation, for an equal mass the abundances present a clear radial
gradient, parallel to the average of the individual linear regressions
derived for each galaxy (solid-line), up to $\sim$2 disk effective
radii. This figure illustrates clearly that the common gradient is
independent of the mass, as mentioned in previous sections.

\begin{figure}
\centering
\includegraphics[width=7.5cm,angle=270]{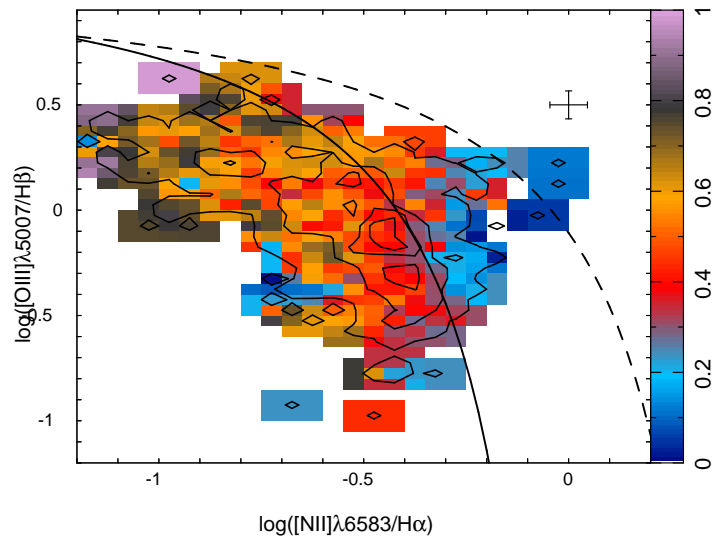}
\caption{\label{fig:OUT_diag} BPT diagnostic diagram similar to the one
 presented in Fig. \ref{fig:BPT}, right-panel, restricted to the \ion{H}{ii}
  regions beyond 2.2 disk effective radius. The contours indicate the number
  density of regions, while the color-coded image represent the fraction of
  young stars ($<$500 Myr). The error bar represents the typical error in both
  parameters for the represented \ion{H}{ii} regions.}
\end{figure}


The right-panel of Fig. \ref{fig:grad_OH}, shows the average
radial distribution of the oxygen abundance after re-scaling to the average
representative abundance. The distribution of \ion{H}{ii} regions is
represented as a density map, both with a color-coded image and a contour map
(with the first contour encircling 95\%) of the regions. The solid-yellow
circles indicate the mean abundances at equal spatial bins of $\Delta
r=$0.3 r$_e$ with the error bars indicating the corresponding
standard deviation. No significant deviation from the monotonic
decrease defined by the average gradient (represented with the black
solid line), is found up to 2.1 disk effective radii. A linear regression
to the scaled distribution of \ion{H}{ii} regions, restricted to the
spatial range between 0.3 and 2.1 r$_e$ (represented by the green
dashed-line) derives a slope of $\alpha_{O/H}-$0.107 dex/$r_e$ and
$\sigma=$0.004 dex/$r_e$, fully compatible with the average value found for
all the sample $\alpha_{O/H}-$0.09 dex/$r_e$ and $\sigma=$0.09 dex/$r_e$
(solid black line).

The origin of this common abundance gradient has to be directly linked to the
formation and evolution of the disk in spiral galaxies.
Recent results, based on the analysis of the star-formation history of CALIFA data,
found undisputed evidence of the inside-out growth of the stellar mass
in galaxies \citep{perez13}, at least for galaxies more massive than
7$\times$10$^{10}$ M$\odot$. Both the extinction-corrected color
gradients in nearby galaxies \citep{mun07}, and weak dependence on the
mass-size relation with redshift \citep{truj04,bard05,truj06} also support
an inside-out scenario for the evolution of disks. 
These results are also supported by the radial distribution of the stellar
ages found for the same data set
(Gonz\'alez Delgado et al., 2013). 

However, the described universal slope for the abundance gradient,
being independent of many of the properties of the galaxies, was only
recently described \citep{sanchez12b}. This result
imposes a more severe restriction to our current understanding of how
disk galaxies grow. In essence, it agrees with the recently proposed
\Sz\ relation \citep{rosales12}, that links the gas abundance with the
mass density of the underlying stellar population. It describes how
the stellar mass and the gas abundance, both fundamental products of
the star formation history, grow side-by-side in disk galaxies, from
the center to the outer-parts. Together with the \mz\ relation, they
indicate that more massive galaxies (that trace the deepest
potential-wells), form before and faster, accumulating more stellar
mass and more metals. The presence of a common gradient in the
abundance indicates that all disk dominated galaxies of the same disk
effective radius (hence, same disk mass, if having similar central
surface brightness) synthesize metals at the same galactocentric
distance with a similar efficiency.

The common slope suggests that the chemical evolution of galaxies is
very similar in all disk galaxies, being compatible with a pseudo closed-box
model. The classical closed-box model considers that each radial bin of
a galaxy comprises primordial gas from which stars are born, live
all their life-time and die {\it in-situ}, according to a given SFR
and IMF prescriptions \citep{pagel75}. Therefore, if the amount of
primordial gas is proportional to the depth of the potential well, and the
efficiency of the SFR is the same for all the galaxies, both the
stellar mass and the enrichment would only be  proportional to the
time, for a given halo mass. The efficiency of the SFR (or
starformation efficiency, SFE), is defined as the SFR surface density per
unit of neutral gas surface density along a line of sight. Recent results
indicate that the SFE does not vary strongly where the ISM is mostly
H$_2$, in spiral galaxies \citep[e.g.][]{leroy08}. Under this
assumption all galaxies should have an universal gradient of their
oxygen abundance with its zero-point proportional to the total mass.

However, it is well known that the closed-box model cannot predict the
right fraction of metal-poor stars with respect to the observed
metallicity distribution of nearby long-lived stars in the Milky Way
\citep[e.g.][]{gibson03}. A more realistic model overcomes this problem
by allowing the disk of galaxies to form via continuous accretion of gas,
driven by the gravitational force (pseudo closed-box model). This
accretion can be compensated or even halted for certain galaxies and
over certain periods by supernova explosions
\citep[e.g.][]{lars74}. However, the outflow of gas is not expected to
feature in the history of most spiral galaxies and is usually
neglected in the models \citep{gibson03}. This modified model is
consistent with the described common radial gradient if the local gas
recycling is faster than other timescales involved \citep[][]{silk93},
and if the radial inflow is similar for those radial bins with the
same stellar mass.

In summary, the radial gradient appears to be the consequence of
different evolutionary rates along the radius. Recent chemical evolution models
suggest that the inner regions evolve faster than the
outer ones \citep{molla05}, following a local downsizing evolution in
agreement with the \Sz\ relation
\cite{rosales12,sanchez13}. Therefore, an evolution in the slope of
the radial gradient expressed in physical scales (dex/kpc) is
expected, with a flattening for the more massive or more evolved
galaxies \citep{molla05}. The described characteristic slope of the radial gradient when normalized to the effective radius (dex/$r_{e}$) implies a tight co-evolution of the radial gradient with the scale-length of the galaxy, with more massive galaxies being larger (which is also observed).

\subsection{The flattening of the abundance gradient in the outer regions}\label{flat}

The linear regressions shown in the right-panel of
Fig. \ref{fig:grad_OH} have been extrapolated beyond r$>$2.2 effective
radii to illustrate the fact that in the outer regions the abundances
present a clear flattening, as already shown in Fig. \ref{fig:grad}
for some individual galaxies. We detected 484 \ion{H}{ii} regions at
these outer areas, corresponding to 94 individual galaxies. In most of
them the derived abundance is larger than the expected by the
extrapolation of the monotonic decrease.

Average radial distributions of oxygen abundances, similar to the one
shown in Fig. \ref{fig:grad_OH}, were created for different subsamples of
galaxies, segregated by their morphology, photometric properties and
the structural parameters. The same subgroups analysed in the previous
section where selected. No significant difference is found in either
the qualitative and quantitative shape and slope of the radial
gradient. It is worth noting that {\em the flattening at the outer regions
  seems to be a universal property of disk galaxies}, independent of the
inclination, mass, luminosity, $C$-index, morphology and/or presence of bars.
Only in the case of the interacting galaxies, with a much lower slope in the abundance
gradient in the inner disk it is unclear whether or not an outer flattening exists (or if all the distribution is flattened). It is also important to
bear in mind that although we have adopted the O3N2-indicator, the
flattening does not depend on the actual strong-line indicator
selected. We tested it using the R23-based calibrator by \cite{tremonti04},
the N2-based calibrator by \cite{pettini04} and the C-method described by
\cite{P12}, with consistent results.

As mentioned before, several previous studies found a similar behavior
for individual or a limited sample of galaxies
\citep[e.g.][]{bresolin09,yoac10,rosales11,mari11,scar11,bresolin12}. We have
found a first hint of this flattening in our pilot analysis of the
face-on galaxies in the CALIFA feasibility sample plus PINGS data, for
a limited set of galaxies and \ion{H}{ii} regions
\citep{sanchez12b}. The same pattern in the abundance was described
(i) in the extended UV disks discovered by GALEX \citep{gil05,thil07};
(ii) in the metallicity gradient of the outer disk of NGC 300 from
single-star CMD analysis \citep{vlaj09}, (iii) in the Milky-way,
based on the spectroscopic analysis of the outer \ion{H}{ii} regions 
\citep{1996MNRAS.280..720V,este13}, using open clusters
\citep[e.g.][]{2008A&A...480...79B,2009A&A...494...95M,2012AJ....144...95Y},
or Cepheids \citep[e.g.][]{2002A&A...392..491A,2003A&A...401..939L,2004A&A...413..159A,2008A&A...490..613L};
and (iv) in the oxygen abundance gradient of different
individual galaxies, like M\,83 \citep{bresolin09}, NGC\,628 \citep{rosales11},
or NGC\,1512 and NGC\,3621 \citep{bresolin12}, to cite just a few
results. However, this is the first unambiguous detection of such a flattening
in most of the galaxies with detected \ion{H}{ii} regions beyond $\sim$2
effective radii.

\begin{figure}
\centering
\includegraphics[width=6.1cm,angle=270,clip,trim=0 0 0 0]{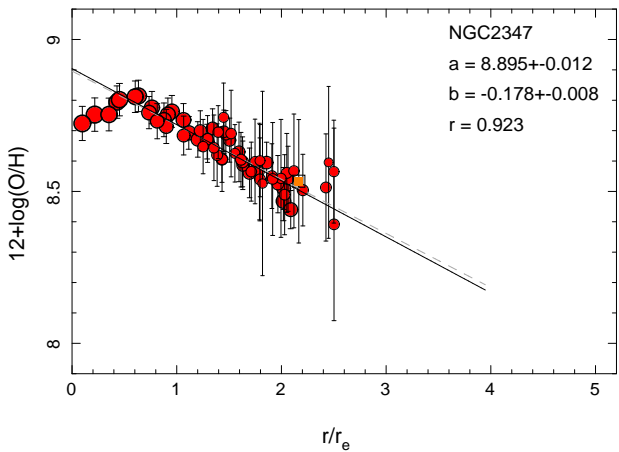}
\caption{\label{fig:drop} Radial distribution of the oxygen abundance derived
  for the individual \ion{H}{ii} regions for NGC\,2347, one of the galaxies
  showing a clear drop of this parameter in the inner regions. The symbols,
  lines and labels are the same as the ones shown in Fig. \ref{fig:grad}.}
\end{figure}

The nature of this flattening is still under debate, in particular
because until the current study it was not even clear if it is a
common feature of all disk-galaxies. The observational and theoretical
investigations of the metal content in the outer regions of galaxies
is relatively recent. Because of their extreme conditions (e.g. very
low gas densities and long dynamical timescales), the outermost
regions of the galaxies are sensitive probes of the mechanisms leading
to the assembly of the disks and their evolution, and thus of great
importance for constraining galactic chemical evolution models.  The
flat abundance distribution in the outer regions of galaxies
corresponds usually to the lowest metallicity values in the
galaxies. Nevertheless, these abundances are still relatively
high. Assuming continuous star formation with the currently observed SFR,
the time required to enrich the ISM up to 12\,+\,log(O/H) $\sim8.2$
has been estimated in $\sim10$ Gyr. \citep{bresolin12}.  According to cosmological
hydrodynamic simulations, in the {\em inside-out} scenario for galaxy
growth the outer disks of galaxies are formed in the last $\sim4-6$
Gyr. \citep[][]{2008MNRAS.389.1137S,2009MNRAS.396..696S}, making
unlikely that the {\em in situ} star formation could have enriched the
interstellar medium to the presently observed values. Therefore, in
the case of isolated galaxies, alternative mechanisms of metal
redistribution must play an important role.

Among the different proposed processes that could produce this
abundance flattening we wish to mention: (i) large-scale processes of angular momentum
transport that produce an intermix of gas abundances, including radial gas
flows \citep[e.g.][]{1985ApJ...290..154L,1992A&A...262..455G,2000A&A...355..929P,2009MNRAS.396..203S,2011A&A...531A..72S},
resonance scattering with transient spiral density waves \citep{2002MNRAS.336..785S},
and the overlap of spiral and bar resonances \citep{2011A&A...527A.147M};
(ii) minor-mergers, and captures of, or perturbations by satellite galaxies that
increase the metal content in the outer regions \citep{2009MNRAS.397.1599Q,2012MNRAS.420..913B}; 
(iii) a non linear Schmidt-Kennicutt law that is able to trigger highly
efficient star-formation process at these distances \citep[e.g.][]{este13}; 
(iv) a spatial association of the flatter area with gas located in the thick disk,
which is known to have different metallicity patterns \citep[e.g.][]{neves09,ishi12};
(v) a plateau of the gas abundance in the intergalactic medium due to the
cosmological evolution of galaxies, and the subsequent pollution of metals
(S\'anchez et al., in prep.); etc.
Stellar radial migration has also been proposed as a possible cause for the
flattening of the metallicity gradients for stellar populations in spiral
galaxies \citep[e.g.][]{2008ApJ...675L..65R,2008ApJ...684L..79R}. However,
this argument is valid only if stars and gas are not decoupled.


The detailed understanding of the nature of the flattening will require
further observational and theoretical efforts. However, we can outline some of
the properties of the \hh regions in this regime, that may help to constrain
and distinguish between the different scenarios summarized before. 
Figure \ref{fig:OUT_diag} shows the BPT diagram for the 484
\ion{H}{ii} regions beyond 2.2 disk effective radii, together with
the fraction of young stars corresponding to each region, as
presented in Fig.\ref{fig:BPT}. Although the number of regions here is more
limited than the whole sample, the distribution as a function of the BPT
diagram is very similar. It shows a clear gradient in the stellar
populations, covering a range of young stars of almost a 100\% to
nearly $\sim$30\%. The range of oxygen abundances measured in these regions is
8.2$<$12\,+\,log(O/H)$<$8.7. Thus, based on the analysed parameters these
regions do not show clear differences with respect to the remaining
population at small galactocentric distances.

As indicated before, even at these galactocentric distances it is
possible to find \ion{H}{ii} regions located above the
\cite{kauffmann03} curve ($\sim$15\% beyond 2.2 effective radii). Due
to their galactocentric distance, these regions cannot be contaminated
by a non-negligible possible central ionizing source (like an AGN),
and the diffuse emission is also extremely weak to affect the line
ratios. A visual inspection of the H$\alpha$ maps reveals that they are
{\it bona fide} \ion{H}{ii} regions, as can be appreciated in
Fig. \ref{fig:HII} and Fig. \ref{fig:grad} for the case of
UGC\,312. This result reinforces our adopted selection criteria,
i.e. by using the information of the underlying stellar population as
a proof for on-going star-formation activity, we do include \hh regions that
would otherwise be discarded by using solely empirical demarcation curves
based on strong emission-lines, like the \cite{kauffmann03}
curve.


\begin{figure}
\centering
\includegraphics[width=6.1cm,angle=270,clip,trim=0 0 0 0]{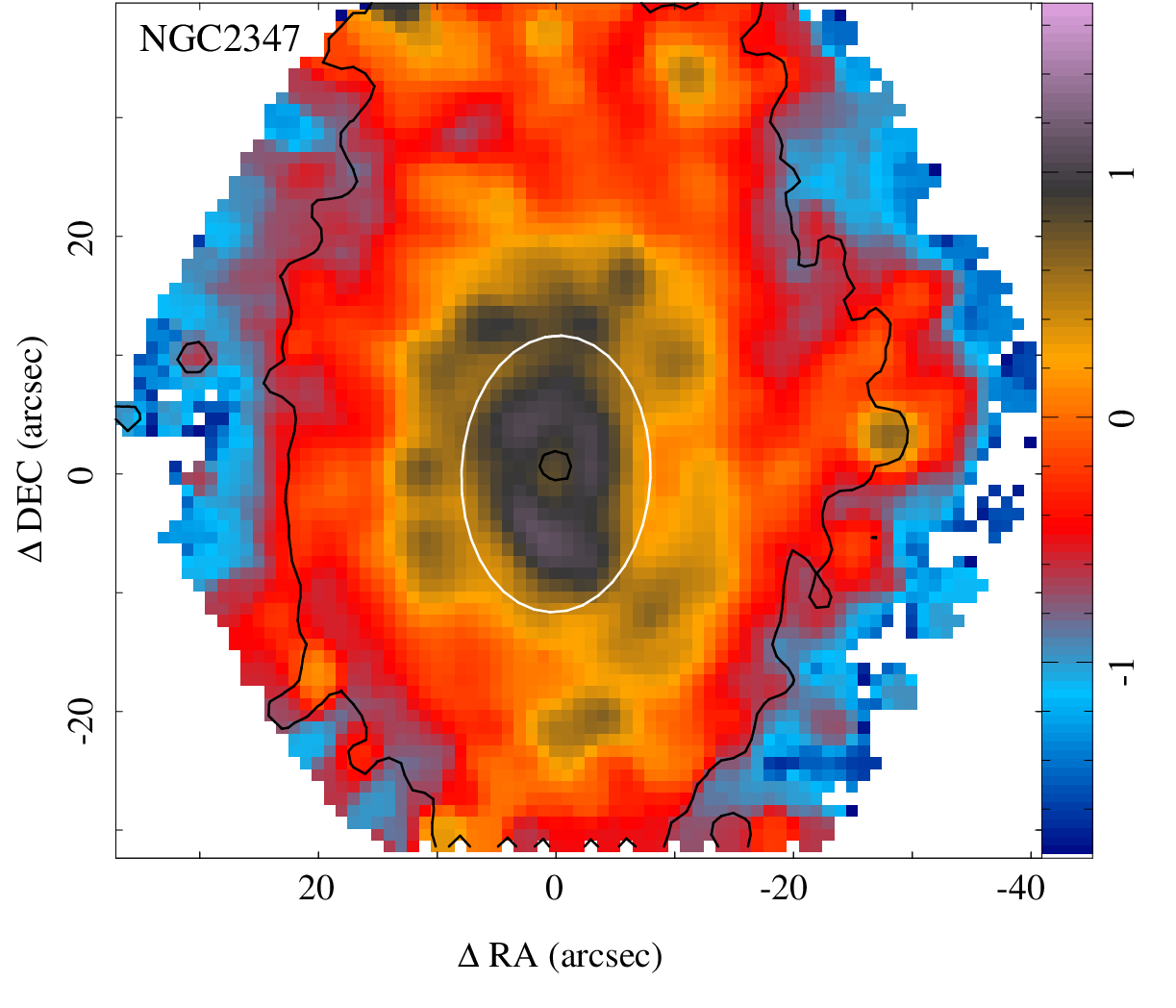}
\caption{\label{fig:drop_Ha} Color-coded image of the H$\alpha$
  intensity flux, in logarithmic scale of \fuDEN units , corresponding to NGC~2347 (the
  galaxy shown in Fig. \ref{fig:drop}). The contours represent the flux
  intensity in the V-band, extracted from the datacube the datacube,
  at two intensity levels, 0.2~\fuden\ (corresponding to $\sim$3
  $\sigma$ detection limit in the continuum) and 15~\fuden, included
  to illustrate the location of the center of the galaxy. The white ellipse indicates the
location of the knee/peak in the radial abundance shown in Fig. \ref{fig:drop}.}
\end{figure}

\subsection{Abundance decrease in the inner regions}\label{drop}

Some galaxies present a truncation in the monotonic increase of the
oxygen abundance towards the center, or even a drop in the inner
regions, at $\sim$0.3-0.5 $r_e$ \citep[e.g.][]{rosales11,sanchez12b}.
In the Milky Way, the metallicity distributions in the Galactic bulge,
in the nuclear bulge and in the Galactic bar display different trends
and are difficult to assess given the obvious geometrical obstacles.
However, iron abundance determinations using stellar tracers (WR,
luminous blue variables, red supergiants) located along the Galactic
bar and in the nuclear bulge are more metal-poor than classical
Cepheids in the inner disk, suggesting that the metallicity
distribution is probably approaching a plateau toward the shortest
Galactocentric distances \citep[$R_G < 5.5$\,kpc,
  {[Fe/H]\,$\sim$0.4}][and references
  therein]{2013A&A...554A.132G}. This flattening has been observed in
other galaxies apart from the Milky Way
\citep[e.g.][]{werk11,bresolin12,sanchez11} and has been predicted by
some chemical evolution models \citep[e.g.][Molla et al., 2013,
  submitted]{molla05}. In some galaxies, such as NGC 628, the decrease
was associated with a starforming ring, with a clear signature in both
the H$\alpha$ emission and the underlying stellar population
\citep[e.g.][]{Wakker:1995p3788,Ganda:2006p3135,fathi07,sanchez11}.
This circumnuclear ring was placed at the expected location of the
inner Lindblad resonance, where gas is expected to accumulate due to
non-circular motions \citep{fathi07}. Despite  this possible
explanation, it is not known if this is a common feature of all
galaxies or, on the contrary, characteristic of a physically distinct
galaxy subclass.

In \cite{sanchez12b}, we found that the drop in the inner region was
present in a substantial fraction of the galaxies, and it was visible
in the average gradient. Figure \ref{fig:grad_OH}, right-panel, shows
the common abundance gradient for our current sample, which comprises
many more galaxies and was properly selected to be statistically
significant (Walcher et al., in prep.). There is no evidence for the a
central drop in the average gradient, indicating that it is not a
common feature of all the galaxies.

We perform a visual inspection of the individual gradient of each
galaxy, classifying them as (i) objects without evidence of an inner
drop in the abundance (146); (ii) objects with a possible evidence of
a drop (21); and (iii) objects with a clear evidence of a drop
(26). Figure \ref{fig:drop} shows the typical abundance gradient of
this latter case, corresponding to NGC\,2347. The galactocentric
distribution shows a monotonic, linear gradient from $\sim$0.6 to
$\sim$2.4 disk effective radius, with a hint of a flattening beyond
this radius, and a clear drop in the central regions ($<$0.6
$r_e$). We should note here that {\it ``drop''} is used as a general term,
defining either a real negative gradient towards the center or just a
flattenning in the inner regions. The number of detected \ion{H}{ii}
regions and the error associated with their corresponding abundances
prevent us from making a firm distinction.

Half of these galaxies are morphologically classified as Sb/Sbc (13 out of 26),
although a KS-test of the morphological distribution compared with the
corresponding one for the full sample (or the sample without evidence
of a drop), indicates that both samples are not significantly
different. The same result is derived when using the galaxies with
possible evidence of an inner drop. Therefore, the drop does not seem
to be associated with a particular morphological type of galaxies.

As we mentioned before, bars may induce radial motions of gas and
stars, which may produce the described effect either by a mix of the
gas content inside the bar length or by a transfer of gas towards the
edge of the bar. Cavichia et al (2013, submitted) have developed a
chemical evolution model in which the bar is taken into account in
driving a radial flow outwards or towards the inner regions. With this
model a slight flattening of the radial gradient of abundances appears
along the disk, but only in the bulge-disk interface region. Thus, at the
corotation radius of the bar there is an increase of SFR and, in
consequence, the oxygen abundance increases.  The net effect is that
there is an apparent decrease or drop of the abundance in the central
regions when actually there is a {\it hump} around 0.5$r_e$. If a bar
were the dominant effect that produces the decrease of abundance in these
galaxies, then it would be expected to detect it more frequently in barred
galaxies. However none of these effects are observed. Although the
sample is still too small to provide a statistical significant
statement, the fraction of barred galaxies within the galaxies with
clear drop of the abundance in the inner areas ($\sim$15\%, 3 out of
21, once excluded the edge on galaxies) is half of the observed for
the total sample ($\sim$28\%, i.e., 53 out of 193).

As we indicated in Sec. \ref{slope_inter}, interactions and mergers
are events that may produce also a change in the distribution of
abundances in galaxies, due to radial motions induced by changes in
the angular momentum patterns. We speculate that these kind of disturbances
may also induce  the detected drop of abundances in the inner regions of
galaxies (e.g. by a minor-merger or a satellite galaxy capture). 
However, when analyzing the interaction stage of our sub-sample of galaxies
with an inner drop we found that none of them presents some evidence of an on-going
interaction/merging event.
Therefore, it seems that this effect is due to the secular evolution
of the galaxy, and not induced, in general, by the presence of an on-going interaction (although it does not discard past interactions or a
minor merger that are not evident in the morphology of the galaxies).

If the origin of the drop is due to radial movements of the gas,
then its presence would also have an effect on the overall distribution of
abundances at larger radii. In order to explore if the presence of an inner
drop has any effect in the determination of the slope of the abundance
gradients (measured between $0.3-2.1$ disk effective radii), we performed the
following test: if we only consider the slopes of the abundance gradients
for the galaxies with clear evidence of a drop, we obtain a mean slope
$\alpha_{O/H,drop}=-$0.15 dex/$r_e$ and $\sigma=$0.06 dex/$r_e$ (26 objects),
which is significantly larger than the slope for the galaxies without evidence
of such a drop, i.e. $\alpha_{O/H,drop}=-$0.08 dex/$r_e$ and $\sigma=$0.08
dex/$r_e$ (146 objects), with a probability of being different of
$p_{KS}=$99.86\%. This is consistent with a radial movement towards the
knee/peak point in the abundance distribution, outwards from the inner regions
and inwards from the outer ones.

A detailed inspection of the H$\alpha$ intensity maps of the less
inclined galaxies shows that in most of them there is a star-forming
ring that is spatially located inside the galactocentric distance
defined by the knee in the radial distribution of the
abundance. Figure \ref{fig:drop_Ha} shows an example of this ring
detected in the H$\alpha$ distribution of NGC\,2347. A similar result
was found for NGC\,628 \citep{sanchez11}, using data with better
physical spatial sampling provided by the PINGS survey
\citep{rosales-ortega10}. We speculate that the presence of the
circular star-formation ring and the drop in the oxygen abundance
inwards (together with the lack of evidence of an interaction and the
low fraction of barred galaxies with these features), suggest that
both features are the consequence of gas radial motions induced by
resonances in the disk-speed pattern \citep{roskar08b}.

We will explore this hypothesis in the future with a detailed analysis of the
stellar and gas kinematics, using the V1200 data available for the CALIFA
objects.

\section{Conclusions}\label{disc}

In this article we described the procedures to detect, select, and
analyse the spectroscopic properties of the largest homogeneous catalog of
\ion{H}{ii} regions and associations  observed  to
date. This catalog comprises more than 5000 ionized regions
associated with star-formation distributed in 227 galaxies of
different morphological types, and evenly distributed along the
color-magnitude diagram (out of 306, that comprised our original
sample). We demonstrated that the combination of the properties of the
ionized gas with the on-going underlying stellar population provides a
robust means for selecting {\it bona fide} \ion{H}{ii} regions,
without excluding regions that would had been disregarded by other
classical methods.

We use strong-line indicators to derived the oxygen abundance, and
the deprojected galactocentric distances to derive the radial
distances of the individual \ion{H}{ii} regions. With this information
we explore the radial abundance gradients for the individual galaxies
in our sample.  The results of this paper show that {\em disk galaxies
  in the local Universe present a common or characteristic gradient in
  the oxygen abundance of $\alpha_{O/H}=-$0.1 dex/$r_e$ up to $\sim$2 disk effective radii}, with a small dispersion compatible with being produced by random fluctuations. No significant differences are found on
the basis of the morphological type, presence or absence of bars,
absolute magnitude and/or stellar mass. A weak trend towards slightly
flatter gradients is found for massive ($>$4.5~10$^{10}$ M$\odot$) and
highly concentrated galaxies ($C>$2.4), although there are galaxies
with few number of \ion{H}{ii} regions and with the less clearly
defined disk. The only clear deviation from the common slope is seen
in galaxies with evidence of interaction or undergoing merging. For these galaxies the gradient is significantly flatter.
Similar results are obtained with other normalization radii than $r_e$, like the
disk scale-length ($r_d$) or the radius at which the surface
brightness reaches 25 mag/arcsec$^2$ ($r_{25}$). The use of
different normalization radii only changes the numerical values of the
common slopes.

These results agree with the main conclusions of our previous study
\citep{sanchez12b}, where a limited sample of 38 face-on spiral
galaxies was analysed, using similar methods as the ones described
here. Both results apparently contradict previous works, in which the
slope of the radial gradients show a trend with certain morphological
characteristics of the galaxies. \cite{VilaCostas:1992p322} and
\cite{zaritsky94} showed that barred spirals have a shallower gradient
than non-barred ones. However, we must recall here that these
statements are based (in general) on gradients constructed on physical/linear
scales (i.e. dex\,kpc$^{-1}$), not on normalized ones (see Appendix\,\ref{GC}
for a discussion on the dependence of the results on the normalization). A
slight decrease in the scatter between the different slopes is found by
\cite{VilaCostas:1992p322} and \cite{diaz89}, when the scale-length is
normalized by the disk-scale, which agrees with our results.
We conjecture that both results would come into agreement if the
size-luminosity/morphological type relation was considered.
In a companion article (S\'anchez-Bl\'azquez, in prep.), we analyse the
radial gradient of stellar metallicity, where we have found consistent
results, i.e. when normalized to the effective radius of the disk, the slope of
the stellar population gradients does not correlate with the mass or with the
morphological type of the galaxies, we do not find any difference in the
metallicity or age gradients in galaxies with and without bars, and the young
stellar population shows a metallicity gradient which is very similar to that
of the gas.

We argue that the common slope suggests that the chemical evolution of
galaxies is very similar in all disk galaxies, being compatible with a
``modified'' closed-box model, in which the disk of galaxies form via
continuous accretion of gas, driven by the gravitational force. In this scenario,
if the amount of primordial gas is proportional to the depth of the potential well, and the
efficiency of the SFR is the same for all the galaxies, both the
stellar mass and the chemical enrichment would be just proportional to the
time, for a given halo mass. Under this assumption all galaxies should
have an universal gradient of their oxygen abundance with its
zero-point proportional to the total mass.
It is important to note that our results do not deny the existence of radial
movements and metal mixing, although they put new constraints on their net effects
on the properties of the galaxies.

Galaxies under interaction or undergoing a merger event show a clearly
flatter oxygen abundance distribution, in agreement with recent
results by \cite{kewley10} and \cite{rich12}. This indicates that
these dynamical processes can produce an effective mixing of metals. On
the contrary, the fact that barred and unbarred galaxies do not
present a clear difference in their abundance slope indicates that
bars either (i) do not significantly enhance the efficiency of chemical mixing; (ii) they
produce a proportional change in the gas abundance and stellar mass
distributions, which compensate each other when normalized by the
disk effective radius or (iii) they are of temporary nature with a
life-time that is shorter than chemical mixing timescale.

Another conclusion reached by this study is that {\em the flattening of the
  abundance gradient at the outer regions seems to be an universal property of
  disk galaxies}. Beyond $\sim$2 disk effective radii, our data shows a clear
evidence of a flat distribution of the oxygen abundance in most of the
galaxies with detected \ion{H}{ii} regions at these radial distances.
This feature has been previously reported by several authors
\citep[e.g.][]{bresolin09,mari11,bresolin12,sanchez12b}, although with less
significant numbers.
Although we cannot provide a conclusive answer regarding the origin of this
flattening, our results suggest that its origin is most probably related to the
secular evolution of galaxies, involving processes like radial migration, or the
capture of evolved satellite galaxies. On the other hand, it disfavors other
scenarios, such as a change in the efficiency of the star formation or the
accreation of chemically polluted intergalactic media. Further analysis is
required to provide a better understanding of this effect.

Finally, we also present observational evidence for a decrease of the
oxygen abundance in the central region of some particular galaxies. Our analysis
indicates that this drop is associated with central star-forming rings.
A plausible explanation would be that both features are produced by
the radial flow of gas induced by resonances in the disk pattern speed.
The slightly increase of the slope beyond the knee/truncation point of
the oxygen abundance, the lack of this feature in interacting systems,
and the fact that it is more frequent in non-barred galaxies support this
scenario.

\begin{acknowledgements}

SFS thanks the director of CEFCA, M. Moles, for his sincere support to
this project.

This study makes uses of the data provided by the Calar Alto Legacy
Integral Field Area (CALIFA) survey (http://califa.caha.es/).

CALIFA is the first legacy survey being performed at Calar Alto. The
CALIFA collaboration would like to thank the IAA-CSIC and MPIA-MPG as
major partners of the observatory, and CAHA itself, for the unique
access to telescope time and support in manpower and infrastructures.
The CALIFA collaboration also thanks the CAHA staff for the dedication
to this project.

Based on observations collected at the Centro Astron\'omico Hispano
Alem\'an (CAHA) at Calar Alto, operated jointly by the
Max-Planck-Institut f\"ur Astronomie and the Instituto de Astrof\'isica de
Andaluc\'ia (CSIC).

We thank the {\it Viabilidad , Dise\~no , Acceso y Mejora } funding program,
ICTS-2009-10, for supporting the initial development of this project.

S.F.S., F.F.R.O. and D. Mast thank the {\it Plan Nacional de Investigaci\'on y
  Desarrollo} funding programs, AYA2010-22111-C03-03 and AYA2010-10904E, of
the Spanish {\it Ministerio de   Ciencia e Innovaci\'on}, for the support
given to this project.

S.F.S thanks the the {\it Ram\'on y Cajal} project RyC-2011-07590 of the
spanish {\it Ministerio de Econom\'\i a y Competitividad}, for the support
giving to this project.

S.F.S. and B.J. acknowledge support  by the grants No. M100031241 and
M100031201 of the Academy of Sciences of the Czech Republic
(ASCR Internal support program of international cooperation projects -
PIPPMS) and by the Czech Republic program for the long-term
development of the research institution No. RVO67985815.

R.G.D , E.P. and R.G.B. thank the {\it Plan Nacional de Investigaci\'on y
  Desarrollo} funding program AYA2010-15081.

F.F.R.O. acknowledges the Mexican National Council for Science and
Technology (CONACYT) for financial support under the programme
Estancias Posdoctorales y Sab\'aticas al Extranjero para la
Consolidaci\'on de Grupos de Investigaci\'on, 2010-2012.

I.M. and J.P. acknowledge financial support from the Spanish grant
AYA2010-15169 and Junta de Andaluc\'{\i}a TIC114 and Excellence Project P08-TIC-03531.

D. M. and A. M.-I. are supported by the Spanish Research Council within
the program JAE-Doc, Junta para la Ampliaci\'on de Estudios, co-funded by
the FSE.

R.A. Marino was also funded by the spanish programme of International Campus
of Excellence Moncloa (CEI).

J. I.-P., J. M. V., A. M.-I. and C. K. have been partially funded by the
projects AYA2010-21887 from the Spanish PNAYA,  CSD2006 - 00070  ``1st Science
with GTC'' from the CONSOLIDER 2010 programme of the Spanish MICINN, and
TIC114 Galaxias y Cosmolog\'{\i}a of the Junta de Andaluc\'{\i}a (Spain). 

M.A.P.T. acknowledges support by the Spanish MICINN through grant
AYA2012-38491-C02-02, and by the Autonomic Government of Andalusia through
grants P08-TIC-4075 and TIC-126.

CJW acknowledges support through the Marie Curie Career Integration
Grant 303912.

Polychronis Papaderos is supported by a Ciencia 2008 contract,
funded by FCT/MCTES (Portugal) and POPH/FSE (EC).

Jean Michel Gomes is supported by grant SFRH/BPD/66958/2009 from FCT (Portugal).

This paper makes use of the Sloan Digital Sky Survey data. Funding for the
SDSS and SDSS-II has been provided by the Alfred P. Sloan Foundation,  the
Participating Institutions,  the National Science Foundation,  the
U.S. Department of Energy,  the National Aeronautics and Space Administration, 
the Japanese Monbukagakusho,  the Max Planck Society,  and the Higher Education
Funding Council for England. The SDSS Web Site is http://www.sdss.org/.

The SDSS is managed by the Astrophysical Research Consortium for the
Participating Institutions. The Participating Institutions are the
American Museum of Natural History,  Astrophysical Institute Potsdam, 
University of Basel,  University of Cambridge,  Case Western Reserve
University,  University of Chicago,  Drexel University,  Fermilab,  the
Institute for Advanced Study,  the Japan Participation Group,  Johns Hopkins
University,  the Joint Institute for Nuclear Astrophysics,  the Kavli
Institute for Particle Astrophysics and Cosmology,  the Korean Scientist
Group,  the Chinese Academy of Sciences (LAMOST),  Los Alamos National
Laboratory,  the Max-Planck-Institute for Astronomy (MPIA),  the
Max-Planck-Institute for Astrophysics (MPA),  New Mexico State University, 
Ohio State University,  University of Pittsburgh,  University of Portsmouth, 
Princeton University,  the United States Naval Observatory,  and the
University of Washington.

This publication makes use of data products from the Two Micron All
Sky Survey, which is a joint project of the University of
Massachusetts and the Infrared Processing and Analysis
Center/California Institute of Technology, funded by the National
Aeronautics and Space Administration and the National Science
Foundation.

\end{acknowledgements}

\bibliography{CALIFAI}
\bibliographystyle{aa}

\appendix

\section{Effective radius of the disk}\label{Re}

\begin{figure*}
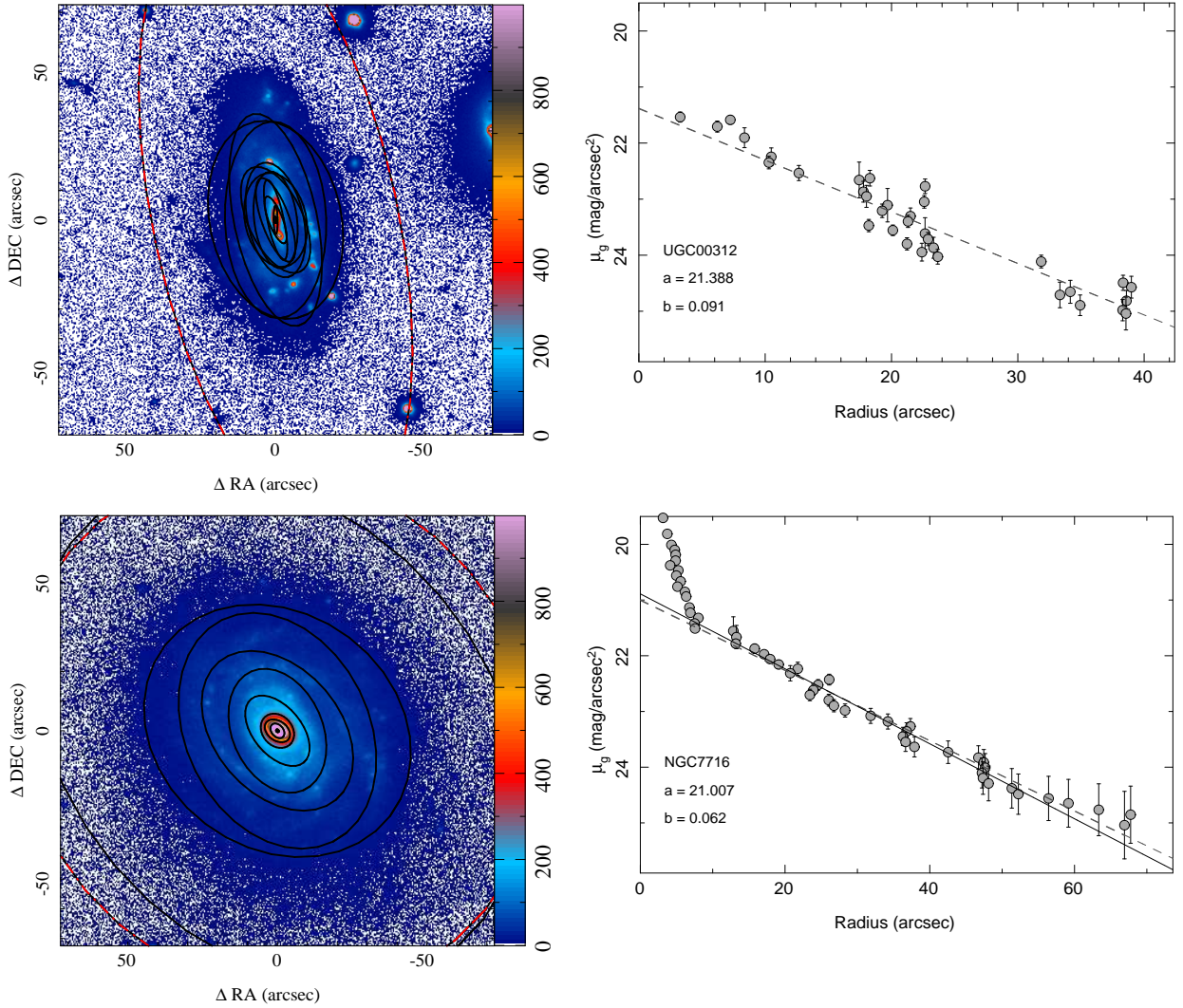

\centering
\includegraphics[width=7.1cm,angle=270,clip,trim=0 0 0 0]{figs/ellipse.UGC00312.NT.ps}
\includegraphics[width=6.1cm,angle=270,clip,trim=0 0 0 0]{figs/94UGC00312.ps}
\includegraphics[width=7.1cm,angle=270,clip,trim=0 0 0 0]{figs/ellipse.NGC7716.NT.ps}
\includegraphics[width=6.1cm,angle=270,clip,trim=0 0 0 0]{figs/94NGC7716.ps}

\caption{\label{fig:iso} {\it Left-panels:} Color-scale representation of 1.5$\arcmin$$\times$1.5$\arcmin$ postage-stamp images extracted from the SDSS $g$-band data (in counts), centred in two CALIFA targets (UGC~00312 and NGC~7716), together with a set of ellipses (solid-black lines) representing the recovered shape at different isophotal intensity levels by the analysis described in the text. The last dashed red-black ellipse represents the 1$\sigma$ isophotal intensity level over the background, adopting the median ellipticity and position angle to plot it. {\it Right-Panel:} Surface brightness profiles derived for the considered galaxies on the basis of the corresponding isophotal analysis (gray solid circles), together with the best fit to an exponential profile for the portion of the surface brightness dominated by the disk (gray dashed line). For NGC~7716 a previous iteration of the fitting procedure is shown,  before the rejection of those values dominated by the bulge (solid-line).   }
\end{figure*}

In \cite{sanchez12b} we showed that the abundance gradient of the
analyzed galaxies presents a common gradient up to $\sim$2 $r_e$, when
normalized to the effective radius of the disk ($r_e$). To repeat the same
analysis for the CALIFA dataset included in this study, in addition to
the abundances of the individual \ion{H}{ii} regions, we need to
derive this structural parameter for each galaxy.

The effective radius of the disk was derived based on an analysis of
the azimuthal surface brightness profile (SBP). To derive the SBP, we
perform an isophotal analysis of the ancillary $g$-band images
collected for the CALIFA galaxies \citep[extracted from the SDSS
  imaging survey,][and Paper\,I]{york00}.

These images were created using the {\tt SDSSmosaic} tool included in
IRAF\footnote{IRAF is distributed by the National Optical Astronomy
  Observatories, which are operated by the Association of Universities
  for Research in Astronomy, Inc., under cooperative agreement with
  the National Science Foundation.} \citep{2009MNRAS.400.1181Z}. {\tt
  SDSSmosaic} takes the galaxy coordinates as input argument and
downloads all the individual SDSS frames and the photometric
information from the SDSS DR7 web site for all 5 bands. These frames are
then stitched together, accurate astrometry is computed and the
photometric calibration is written into the header. After that,
the background is subtracted from each scan by fitting a plane surface
(allowing for linear gradient along the scan direction, constant in
the perpendicular direction) outside a circle centered on the
source. Finally, stripes are combined in a mosaic using the program
{\tt Swarp} \citep{2002ASPC..281..228B}. The final mosaic contains the
photometric zero point (P\_ZP keyword) in mag per second of exposure
time (EXPTIME keyword). We extracted postage-stamp images of
3$\arcmin$$\times$3$\arcmin$ size, centred in the CALIFA targets, for
the $g$-band mosaic images to derive the disk effective radius.

\begin{figure*}
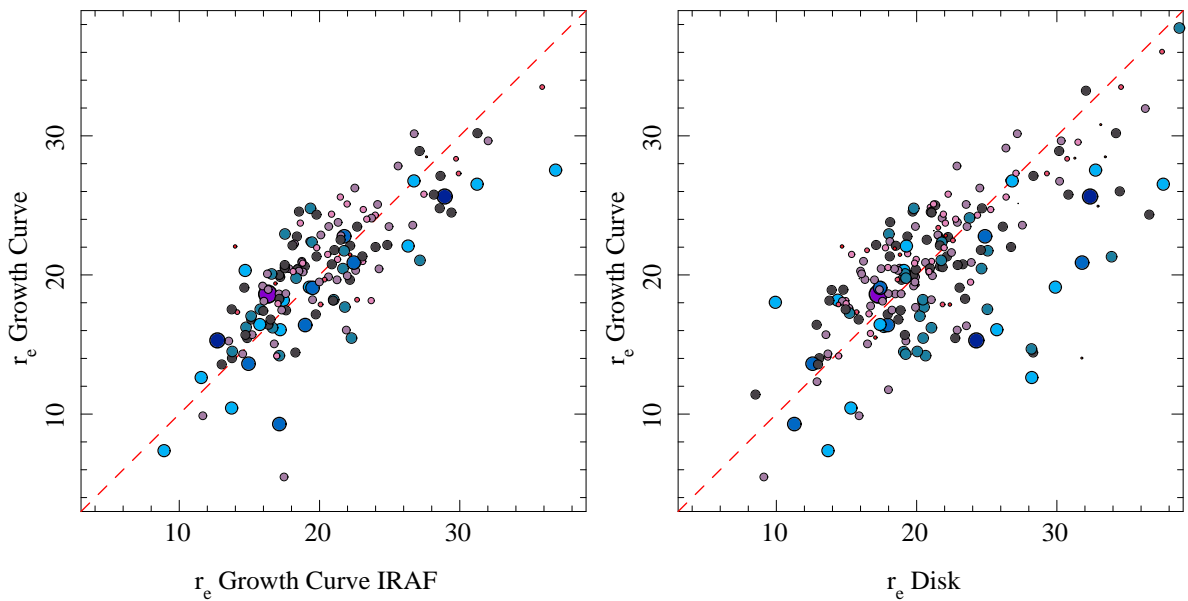

\centering
\includegraphics[width=8.1cm,angle=270,clip,trim=0 0 0 0]{figs/comp_Re_Lucie_GC.ps}
\includegraphics[width=8.1cm,angle=270,clip,trim=0 0 0 0]{figs/comp_Re_Disk_GC.ps}
\caption{\label{fig:comp_Re} {\it Left panel:} Distribution of the effective radius of the analysed galaxies, derived from a growth curve analysis (i.e., the total light effective radius), as a function of the same effective radius derived by a similar growth curve analysis using the surface-brightness profiles derived using the {\tt ellipse} task in IRAF. {\it Right panel:} Distribution of the effective radius of the analysed galaxies, derived from a growth curve analysis (i.e., the total light effective radius), as a function of the effective radius of the disk, derived from the exponential fitting to the surface brightness profile. In both panels, larger and bluer symbols correspond to earlier-type galaxies, while smaller and pinky/reddish ones correspond to later-type galaxies.}
\end{figure*}

The isophotal analysis was performed using the {\tt
  ellipse\_isophot\_seg.pl} tool included in the \textsc{HIIexplorer}
package\footnote{\url{http://www.caha.es/sanchez/HII_explorer/}}).
Contrary to other tools, like {\tt ellipse} included in IRAF, this
tool does not assume \emph{a priori} a certain parametric shape for the
isophotal distributions. The following procedures were followed for
each postage-stamp image: (i) the peak intensity emission within a certain
distance of a user defined center of the galaxy was derived. Then,
any region around a peak emission above a certain fraction of the
galaxy intensity peak is masked, which effectively masks the brightest
foreground stars; (ii) once  the peak intensity is derived, the image is
segmented in consecutive levels following a logarithmic scale from
this peak value, using the equation:

\begin{equation}
{\rm seg(i,j) = ~n_{levels} \frac{log_{10}~flux(i,j)}{log_{10}~flux_{peak} } }
\end{equation}

where {\it seg(i,j)} is the final segmented index at the pixel {\it
  (i,j)}, being an integer number; {\it n$_{levels}$} is the number of
selected levels of the isophotal analysis (100 in our case); {\it
  flux(i,j)} is the intensity at the corresponding pixel {\it (i,j)}
and {\it flux$_{peak}$} is the peak intensity at the center of the
galaxy; (iii) once the image is segmented in $n_{levels}$ isophotal
regions, for each of them, a set of structural
parameters was derived, including the mean flux intensity and the corresponding
standard deviation, the semi-major and semi-minor axis lengths, the
ellipticity, the position-angle and the barycenter coordinates. These
parameters are intended to describe each isophote following an
elliptical shape, but without performing direct fit, which is in
principle more stable in lower signal-to-noise regimes. Figure
\ref{fig:iso}, left panels, illustrates the process, showing for two
particular objects a central section of
1.5$\arcmin$$\times$1.5$\arcmin$ of the postage-stamp images used in this
analysis, together with a set of ellipses plotted using the recovered
shape parameters for some particular isophotes.

{The isophotal segmentation (2nd step of the described procedure), was
first introduced by \cite{papa02}. \cite{noes03} and \cite{noes06}
showed that the surface-brightness profiles derived using this
technique were very similar to those derived using more broadly
adopted techniques, like the 2D GALFIT code \citep{galfit}. We have
adopted this isophotal annuli procedure in previous studies, e.g.\
\cite{kehrig12}, \cite{papa13} and Mast et al. (submitted). The
improvement with respect to a pure isophotal segmentation was the
additional derivation of the structural parameters for each isophotal
region, described before (step iii). }

The isophotal analysis provides  a  SBP, which is then
analyzed to derive the required effective radius. Figure
\ref{fig:iso}, right panels, show two examples of the derived SBPs for
those galaxies shown in the left panels. In \citep{sanchez12b}, where
all the galaxies were clearly disk-dominated, the profiles were
fitted using a pure exponential profile, following the classical
formula,

\begin{equation}
I = I_0 \exp[ - (r/r_d)]
\end{equation}

\noindent
where $I_0$ is the central intensity, and $r_d$ is the disk
scale-length \citep{free70}, using a simple polynomial regression
fitting. The scale-length was then used to derive the disk effective
radius, defined as the radius at which the integrated flux is half of
the total one for a disk component, by integrating the previous
formula, and deriving the relation:

\begin{equation}
r_e = 1.67835 r_d 
\end{equation}

\noindent In the current study, the sample comprises galaxies of
different morphological types.  In many cases the presence of a bulge
prevents us from doing a direct exponential fitting for the complete
SBP. In those cases the effective radius of the disk diverge from the
effective radius of the complete galaxy, defined as the semi-major
axis encircling half of the light of the galaxy. 

First, in order to illustrate that our isophotal analysis provides
consistent results with more classical analysis tools, we determined the
effective radius using a classical growth-curve analysis, based both
on our procedure and the surface-brightness analysis provided by the
{\tt ellipse} task included in IRAF. This latter analysis is part of
the CALIFA sample characterization effort, that can be performed only
in regular shaped targets (Walcher et al., in prep.). Figure
\ref{fig:comp_Re}, left panel shows the comparison between both
effective radii. Different sizes and colors are used to illustrate
the morphological type of the galaxies, with smaller and pink symbols
corresponding to later type galaxies, and bluer and larger symbols
corresponding to earlier type ones. There is an almost one-to-one
relation between both parameters, without any evident difference
between different morphological types. The comparison of other shape
parameters, like the position angle or the ellipticity produce similar
results.

\begin{figure*}
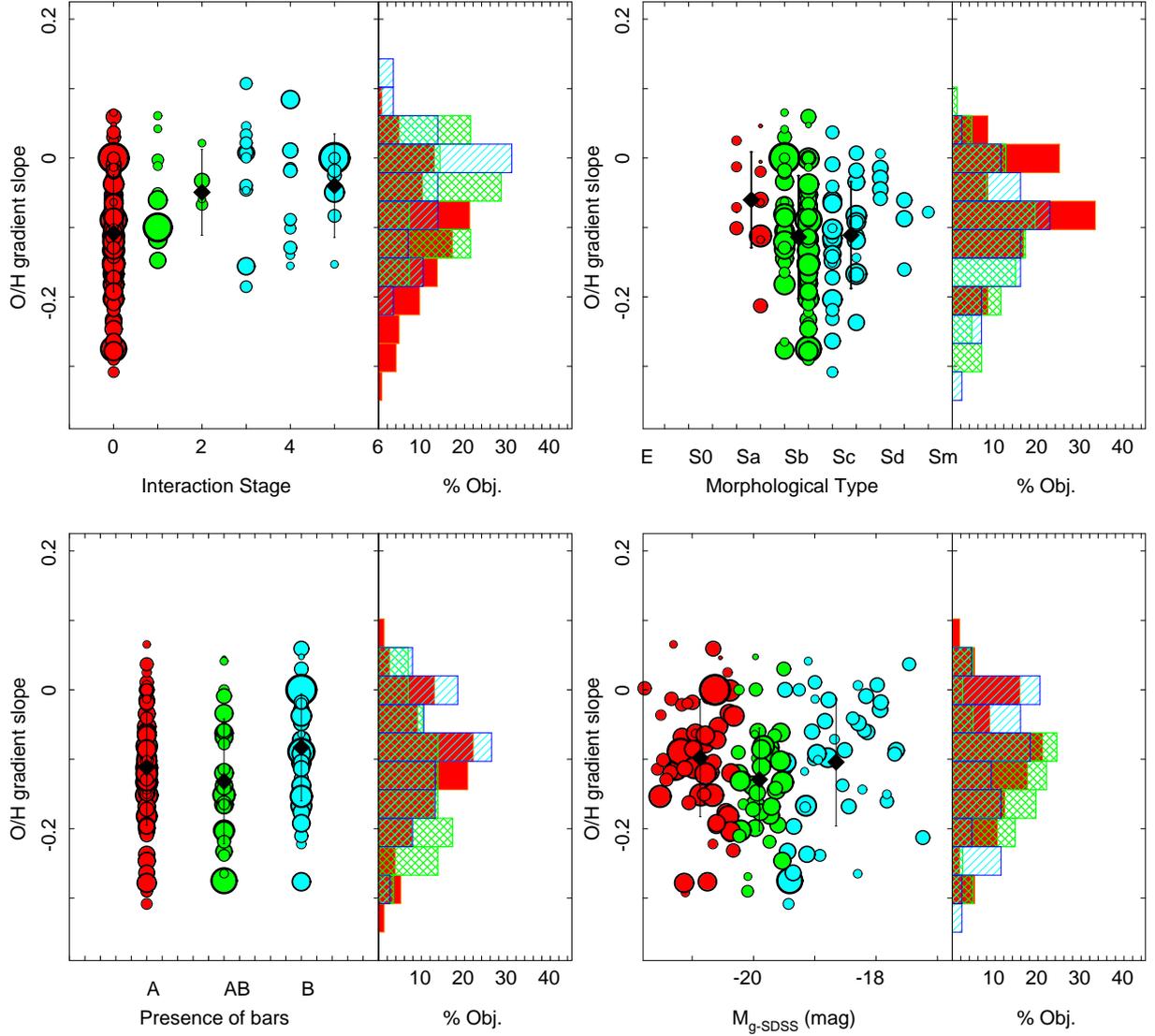

\centering
\includegraphics[width=7.5cm,angle=270,clip,trim=0 0 0 0]{figs/M6Mass_OH_3_GC.ps}\includegraphics[width=7.5cm,angle=270,clip,trim=0 0 0 0]{figs/6Mass_OH_no_inter_GC.ps}
\includegraphics[width=7.5cm,angle=270,clip,trim=0 0 0 0]{figs/16Mass_OH_no_inter_GC.ps}\includegraphics[width=7.5cm,angle=270,clip,trim=0 0 0 0]{figs/36Mass_OH_no_inter_GC.ps}
\caption{\label{fig:morph_GC}{\it Top-left panel:} Distribution of the slopes of the abundance gradients as a function of the interaction stage of the galaxies, when the galactocentric distances are normalized to the effective radius derived by a growth curve analysis. The colors of the symbols and the corresponding histograms indicate three types of galaxies based on the interaction: (i) no signatures of interaction (red), (ii) galaxies with close companions and/or in an early interaction stage (green), and (iii) galaxies under clear collision or evolved mergers (blue). {\it Top-right panel:} Similar distribution of slopes as a function of the morphological classification of the galaxies. The colors of the symbols and the corresponding histograms indicate three types of galaxies based on their morphology: (i) Early spirals, SO-Sa (red), (ii) intermediate spirals, Sab-Sb (green) and (iii) Late spirals, Sc-Sm (blue).  {\it Bottom-left panel:} Similar distribution of slopes as a function of the presence or absence of bars. The colors of the symbols and the corresponding histograms indicate three types of galaxies: (i) clearly non-barred (red), (ii) not clear if there is a bar or not (green) and (iii) clearly barred galaxies (blue). {\it Bottom-right panel:} Similar distribution of slopes as a function of the absolute magnitude of the galaxies. The colors of the symbols and the corresponding histograms indicate three types of galaxies based on the luminosity: (i) luminous galaxies, $M_{g-SDSS}<-20.25$ mag (red), (ii) intermediate galaxies $-19.5<M_{g-SDSS}<-20.25$ mag (green) and (iii) faint galaxies $M_{g-SDSS}>-19.5$ mag (blue). The size of the symbols are inversely proportional to the derived error in the slope for all the panels. }
\end{figure*}

To derive the disk effective radius it is required to fit that the outer portion
of the SBP clearly dominated by an exponential disk. As  is
illustrated in the right panels of Fig. \ref{fig:iso}, in some cases
there is almost no deviation from an exponential disk over  the full spatial
range covered by the SBP (e.g., UGC00312).  However, in other cases
there is a clear deviation in the inner regions due to the presence of
a bulge (e.g., NGC7716). To minimize the effect of the bulge in this
derivation we perform an iterative procedure, in which the SBP,
represented in surface brightness magnitudes, was fitted with a linear
regression. In each stage of the iteration the brightest value of the
SBP was removed, and the regression was repeated. The iteration stops
when only half of the original values remains in the SBP. From the 
set of regressions it was adopted that one with the highest correlation
coefficient between the semi-major axis and the surface brightness magnitude.

The procedure was tested visually, as illustrated in
Fig. \ref{fig:iso}, showing that it provides  a good fit for the
linear regime (i.e., the disk-dominated regime), excluding in most of
the cases the central regions dominated by a bulge. Obviously, the procedure
works better for those galaxies that are still dominated by disk in most of the
SBP, and provides the worst results for those ones dominated by a bulge. However,
this would be a general limitation to any other proposed method with the same aim.

Figure \ref{fig:comp_Re}, right panel, shows the comparison between
the effective radius derived using the growth-curve method and the the
disk effective radius extracted from the iterative fit of the SBP. The
size and color of the symbols represent the same morphological
segregation shown in the left panel. As expected the late-type
galaxies, disk-dominated, are clustered closer to the one-to-one
relation, while most of the earlier-type ones, with brighter bulges,
are shifted towards larger disk effective radius.

\section{Dependence of the results on the derivation of the disk effective radius}\label{GC}

The main result described along the current study is that all
undisturbed galaxies with a disk present a similar radial abundance
gradient, with a characteristic slope, when the galactocentric
distances are normalized to the disk effective radius. This
characteristic slope seems to be independent of other properties of
the galaxies, like morphological types, presence of a bars and
absolute magnitudes or stellar masses. However, this result relies on
the definition of the disk effective radius and its derivation, based
on the surface brightness profile analysis described in the previous
section.

The disk effective radius is a non-standard scale-length, that we have
introduced to characterize the size of the disk in galaxies with
different morphologies. Therefore, it is important to illustrate how
our results are affected by this adopted normalization of the radial
distances. To do so we repeat the analysis using the standard
effective radius derived using the growth-curve analysis ($r_{e,GC}$), to
normalize the abundance gradients, instead of the disk effective
radius. As we already indicated the growth curve effective radius has
been derived independently using to different procedures providing
reliable and consistent results (Fig. \ref{fig:comp_Re}, left panel).

Figure \ref{fig:morph_GC} shows the distribution of the new slopes
obtained when normalizing to these effective radius, along the same
structural parameters of the galaxies shown in Fig. \ref{fig:morph}.
Despite  the fact that individual slopes change, in particular for
the galaxies of earlier type, all the results described in Section
\ref{morph} remain valid:

\begin{itemize}

\item Galaxies with evidence of interaction or in a clear merging
  process present an oxygen abundance gradient flatter than those
  without any clear evidence of interaction. The difference is
  statistically significant, both using KS- or a $t$-test for the
  different distributions. The characteristic slope for non-interacting galaxies it is not affected by the selection of the
  normalization radius, with a mean value of
  $\alpha_{O/H}=-$0.11$\pm$0.09 dex/$r_e$ (although the dispersion
  suffer a slightly increase).

\item The average slope for earlier type galaxies is slightly flatter
  when normalizing by $r_{e,GC}$, instead of the disk $r_{e}$:
  $\alpha_{O/H,Sa/S0}=-$0.06$\pm$0.08 dex/$r_{e,GC}$ instead of
  $\alpha_{O/H,Sa/S0}=-$0.08$\pm$0.08 dex/$r_{e}$. However, there is no
  significant different in the distribution of slopes, either using a
  KS- or a $t$-test analysis.

\item Neither the average slopes nor the distribution of slopes change
  significantly depending on the presence or absence of a bar in the
  galaxies.  The use of the total or disk effective radius seems to be
  irrelevant for the comparison of abundance gradients for barred an
  unbarred galaxies.

\item The abundance gradient slopes normalized by $r_{e,GC}$ do not
  present any dependence in the absolute magnitudes, the stellar
  masses or the concentration indices of the galaxies.

\end{itemize}

In summary, although the slopes of the individual gradients for each
galaxy change when normalizing by the disk $r_{e}$ or the $r_{e,GC}$,
in general the statistical results are basically the same. The main
effect, as expected, is found in the slope of the earlier type
galaxies (Sa/S0), that present slightly flatter gradients. This is
expected, since for these galaxies the disk $r_{e}$ is larger than the
growth curve one, due to the presence of a bulge. For those galaxies
the derivation of the disk $r_{e}$ is also more complicated, for the
same reason. However, the fact that the slope of the abundance
gradient becomes more similar for these galaxies to the one derived
for the latter type ones, when using the disk effective radius,
indicate that (i) the use of this radius provides a better
characterization for the gradient and (ii) the metal enrichment seems
to be clearly dominated by the growth of the disk, rather than other
non-secular processes.

\end{document}